\documentclass[journal=jacsat,manuscript=article]{achemso}
\SectionNumbersOn

\usepackage[version=3]{mhchem} 
\usepackage{siunitx}
\usepackage{booktabs} 

\setkeys{acs}{articletitle=true}

\author{Moritz Bensberg\footnote{ORCID: 0000-0002-3479-4772}}
\affiliation
{Theoretische Organische Chemie, Organisch-Chemisches Institut, Center for Multiscale Theory and Computation, Westf\"alische Wilhelms-Universit\"at M\"unster, Corrensstraße 36, 48149 M\"unster, Germany}
\author{Paul L. T\"urtscher}
\affiliation{ETH Z\"urich, Laboratorium f\"ur Physikalische Chemie, Vladimir-Prelog-Weg 2, 8093 Z\"urich, Switzerland}
\author{Jan P. Unsleber\footnote{ORCID: 0000-0003-3465-5788}}
\affiliation{ETH Z\"urich, Laboratorium f\"ur Physikalische Chemie, Vladimir-Prelog-Weg 2, 8093 Z\"urich, Switzerland}
\author{Markus Reiher\footnote{ORCID: 0000-0002-9508-1565}}
\affiliation{ETH Z\"urich, Laboratorium f\"ur Physikalische Chemie, Vladimir-Prelog-Weg 2, 8093 Z\"urich, Switzerland}
\email{markus.reiher@phys.chem.ethz.ch}
\author{Johannes Neugebauer\footnote{ORCID: 0000-0002-8923-4684}}
\affiliation
{Theoretische Organische Chemie, Organisch-Chemisches Institut, Center for Multiscale Theory and Computation, Westf\"alische Wilhelms-Universit\"at M\"unster, Corrensstraße 36, 48149 M\"unster, Germany}
\email{j.neugebauer@uni-muenster.de}

\title{Solvation Free Energies in Subsystem Density Functional Theory}


\begin{document}


\vspace{3cm}
Date: 25 August 2021

\newpage
\begin{abstract}
For many chemical processes the accurate description of solvent effects are vitally important.
Here, we describe a hybrid ansatz for the explicit quantum mechanical description of solute--solvent and 
solvent--solvent interactions based on subsystem density functional theory and continuum solvation schemes.
Since explicit solvent molecules may compromise scalability of the model and transferability of the predicted solvent effect,
we aim to retain both,
for different solutes as well as for different solvents.
The key for the transferability is the consistent subsystem decomposition of solute and solvent.
The key for the scalability is the performance of subsystem DFT for increasing numbers of subsystems.
We investigate molecular dynamics and
stationary point sampling of solvent configurations and compare the resulting (Gibbs) free energies
to experiment and theoretical methods.
We can show that with our hybrid model reaction barriers and reaction energies are accurately reproduced compared to experimental data.
\end{abstract}

\section{Introduction}
Solvation effects can strongly affect properties of molecular systems.\cite{Kirchner2007,Reichardt2010}
Accordingly, it is decisive to describe these effects accurately when modeling chemical systems in solution.
While it is possible that a solute may react with the solvent, even molecular systems where such reactions do not occur can be significantly modulated by the solvent.
For the calculation of solvation free energies
four main strategies are employed: (i) The calculation of solvation free
energies for an explicit solvation model, \emph{e.g.}, in molecular dynamics simulations. These simulations are often
based on free energy perturbation theory (FEP)\cite{Jorgensen1983, Acevedo2014} and require an extensive sampling of solvent
configurations. This limits the energy functionals used in FEP calculations to classical force fields or
semi-empirical methods.
(ii) The calculation of solvation free energies is also possible in continuum solvation models such as the polarizable continuum
model (PCM)\cite{tomasi2005quantum,Mennucci2012,Herbert2021}, the conductor‐like screening model (COSMO), and its variant
for realistic solvation (COSMO-RS)\cite{Klamt2017}. In these models,
the evaluation of all energy terms can be reduced to an integration over a
molecular cavity of the solute. No explicit averaging over solvent configurations is required, which makes these models
computationally very efficient. However, strongly directional interactions such as hydrogen bonding may
not be modelled accurately. This can be resolved in so-called cluster-continuum models\cite{Pliego2019}.
(iii) These cluster-continuum models consider a few solvent molecules explicitly to overcome
the deficiencies of continuum solvation models and describe the bulk of the solvent as a dielectric continuum\cite{Sicinska2002,Silva2003,Daver2018}. 
For these models, the free energy of solvation can be calculated
in a thermodynamic cycle\cite{Pliego2001,Rempe2000,Sure2021} or by linear response theory\cite{Lima2010,Lima2011}.
In general, it is necessary to sample the configurations of the explicitly considered solvent
molecules for large solvation shells because of the network of non-covalent interactions.
Such sampling may rest on structures taken from molecular dynamics
simulations\cite{Lima2010,Lima2011,Steiner2021} or by optimizing clusters starting from different initial
configurations\cite{Grabowski2002,Silva2009,Florez2018} and weighting the geometries by a Boltzmann distribution\cite{Simm2020}.

(iv) The fourth approach is based on machine learning models, which were shown
to predict solvation free energies for a wide range of molecular
systems\cite{Lim2019,Weinreich2021} with a similar accuracy as
continuum-solvation models since reference data for solvation free energies are
readily available, \emph{e.g.}, through the \textsc{FreeSolv}
data-base\cite{Matos2017}.

The two major challenges for cluster-continuum models are the sampling of configurations of the explicitly considered solvent molecules and the definition of the functional for the solvation free energy.
For feasibility reasons, the configurational sampling of explicit solvent molecules is often limited, by the unfavorable scaling of the quantum chemical methods with system size.

In this work, we therefore focus on the application of subsystem density functional theory (sDFT) to tackle this problem.
The favorable scaling of sDFT with the number of explicit solvent molecules allows us to describe large solvent clusters and a large number of configurations.
In the following section, we elaborate on requirements and advantages of our ansatz and derive the expressions used for the solvation free energy in the case of an sDFT-hybrid solvation model.
We strike a balance between the computational cost of the solvent description and the transferability of results by means of large enough solvent samples.
At the same time, we still rely on a quantum mechanical description of the interactions between solvent and solute for the close vicinity of the solute.
Given the dependence of the results on the sampling of explicit solvent molecule configurations, we scrutinize the new sDFT-hybrid model by two sampling methods (molecular dynamics and random minimum structures) by
comparing to experimental data and pure continuum solvation results.

\section{Theory}
\subsection{The Scaling Advantage of Subsystem Density Functional Theory}
The first attractive property of sDFT is its scaling with the increase in the number of solvent molecules,
\emph{i.e.}, with the number of subsystems in the sDFT calculation, and therefore the size of the explicitly considered
molecular system. Formally, sDFT scales quadratically with the number subsystem if no
approximations to the Coulomb interaction are employed and does not introduce any significant overhead\cite{Jacob2014}. 
In contrast, common Kohn--Sham density functional theory (KS-DFT) formally shows cubic scaling
with system size for three dimensional systems or even a less favorable scaling if
the calculation of exact exchange is required for the chosen functional.

The scaling of sDFT can even be improved if the electron density of a part of the system (\emph{e.g.}, the electron density of very distant solvent molecules) is not optimized during the calculation, which makes
it practically equivalent to frozen density embedding (FDE)\cite{Wesolowski2015}. FDE shows linear scaling
with the number of ``frozen'' subsystems. 
In fact, for the three-partition FDE method\cite{Jacob2008} using the COSMO implicit solvation model, the calculation of the COSMO contribution became the computationally most demanding step, which led to the development of the local-COSMO (LoCOSMO)\cite{Goez2015} approximation.

Furthermore, the subsystem nature of sDFT can be easily exploited for the additional acceleration of the calculations, especially in the case of solvation.
Here, the solvent molecules will always have a very similar molecular structure and, on average, also similar electron density distributions. Hence, it will be sufficient for an initial guess to calculate the electron density of a single solvent molecule and duplicate it to the remaining molecules.
Accordingly, generating an average solvent molecule and skipping the orbital optimizations of the solvent molecules all together can be 
a trade-off between speed and accuracy.

\subsection{The Transferability Advantage of Subsystem Density Functional Theory}
The second reason for choosing sDFT over other supersystem-type methods
is the transferability of the resulting subsystem properties of the solute.
Transferable results are of key interest, for instance, for automated reaction exploration techniques\cite{Unsleber2020}.
Automatizing the screening of solvent effects and the possibility to reuse archived data are clearly beneficial.

Note, that for all considerations in the following, we assume that the solvent is innocent.
This means that if reactions are discussed, all molecules that are part of the reaction are considered as part of the solute; in terms of sDFT they constitute a single active subsystem.
Hence, it is possible to describe reactions involving solvent-type molecules, but for the sake of our nomenclature in this work they would not be considered solvent anymore.
A heuristic for the automatic detection of solvent molecules that should be added to the solute system (active system) is certainly of interest, but is not explored within this work.
We differentiate between two distinct types of transferability: solute transferability and solvent transferability.\\
\\
Solute transferability shall be a measure of the possibility to transfer and compare data and properties that are calculated for
specific solutes.
Within subsystem density functional theory it is possible to couple environmental responses back into the solute by means of the subsystem ansatz.\cite{neugebauer2007}
Hence, sDFT has the key advantage that it generates subsystem properties and data by default, but does capture quantum mechanical environment effects.
It was employed in several studies to model solvent effects on electronic excitations\cite{Neugebauer2005,Neugebauer2005a,Kaminski2010}. 
Assuming that the same sDFT description (\textit{e.g.}, basis set and functionals) is applied, calculated subsystem properties drawn directly from sDFT are directly transferable.
It is inherently apparent which data are integrated from the active system (here: solute) and which are integrated across the environment (here: solvent).
By contrast, data taken from supersystem calculations usually requires an additional localization step that introduces additional ambiguity.
Specifically, different localization techniques such as orbital localization\cite{Lowdin1955,Foster1960,Edmiston1963,Knizia2013} and energy decomposition analyses\cite{Kitaura1976,ziegler1977calculation,Bickelhaupt2003} have
different definitions of the same system and might not be easily transferable to other solutes.

In fact, this interpretation of sDFT as a form of constraint DFT was exploited by Warshel and coworkers to validate and parameterize their empirical valence bond (EVB) method\cite{Hong2006,Xiang2008} for the calculation of reaction free energies in solution.
Furthermore, they calculated solvation free energies with FEP by transferring from the EVB potential energy surface to the sDFT surface\cite{Wesolowski1996a, wesolowski1994ab}.\\
\\
Solvent transferability describes the possibility to transfer and compare data and properties that are calculated for different solvents.
In PCM for example, a calculation of properties
of a single solute in two different solvents can be compared directly, and hence, perfect solvent transferability is guaranteed.
With the ansatz described in this work, we aim at generating a procedure with this property.
Because we want to add some form of explicit description of the solvent, our sDFT approach requires sampling of multiple explicit solvent configurations.
Furthermore, we require convergence with the number of explicitly described solvent molecules.
Only then it is possible to compare solvated structures consistently.
Given that both sampling and the convergence of the explicit solvent are expected to be computationally demanding, sDFT can again be the key ansatz to make these steps more feasible.

\subsection{Reference States}
\begin{figure}[ht]
  \centering
  \includegraphics[width=0.4\textwidth]{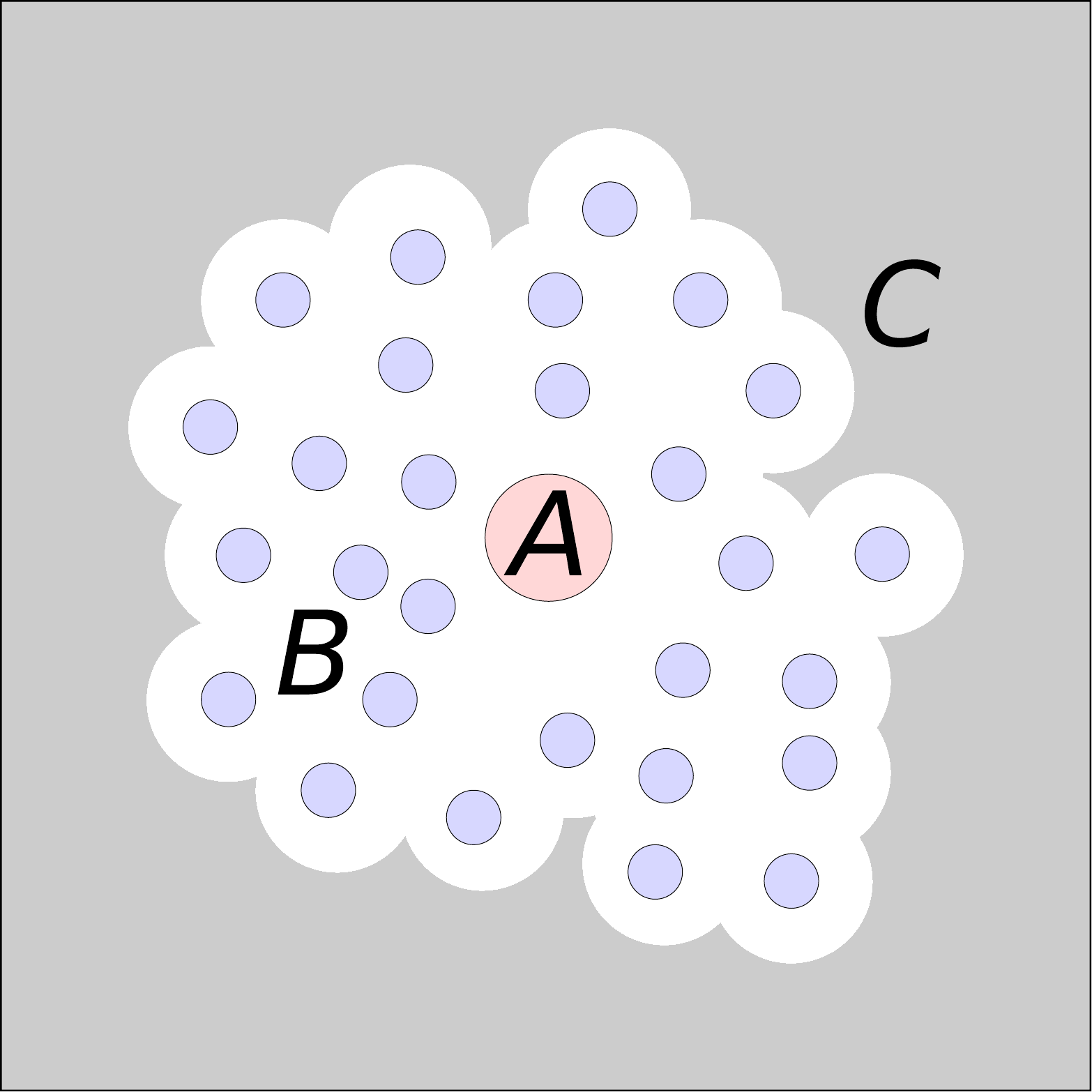}
  \caption{Schematic representation of the sDFT-hybrid model. A solute ($A$, red) is embedded in 
           a set of subsystems ($B$, blue) which serve as an
           explicit representation of the continuum. The cluster
           is then embedded in a continuum model ($C$, contour
           line).}
  \label{fig:sDFTPlusPCM}
\end{figure}

We now describe the definition of the free energy of solvation in a hybrid sDFT--PCM approach.
The effects acting on a solute molecule are described by explicitly
including solvent molecules in this sDFT-hybrid model,
as illustrated in Fig.~\ref{fig:sDFTPlusPCM}.

A key goal of our hybrid solvation approach is to provide the change of free energy of solvation
$\Delta G_\mathrm{solvation}$ for the change of a well-defined reference state for solvent and
solute to its final state as an interacting cluster. A first task is
the definition of the initial reference state.

For the PCM this state is defined as the isolated solute molecule and a continuum solvent where no local
charges are present\cite{tomasi2005quantum}. Such a solvent does not show
electrostatic interactions in this initial state with any external charge and must be polarized first.
For sDFT, such a reference state was not discussed before in literature. However, if such a state is expressed with a single solvent and solute structure, it will in general
show local charges and interact with a external charge.
For the solute the isolated molecule appears to be the natural reference. However, since we are
interested in the combination of sDFT and PCM it is important that we try to choose the same or at least a
very similar reference for the continuum and the explicit solvent. Since the PCM is a well-established model, we reference the unpolarized continuum with sDFT.

\begin{figure}[ht]
  \centering
  \includegraphics[width=0.95\textwidth]{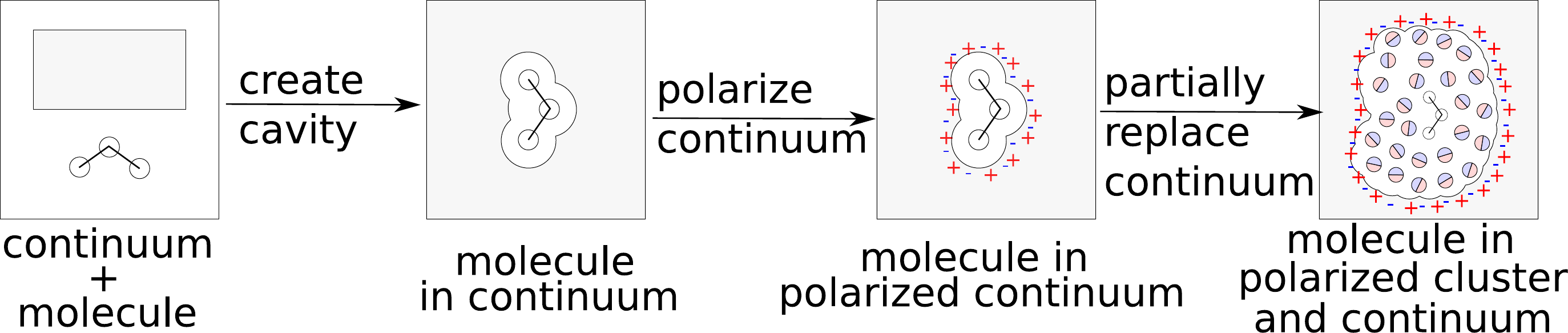}
  \caption{Theoretical model process: Solvation process starting from the unpolarized continuum solvent, inserting the molecule
           into the solvent by creating a cavity, and polarizing the solvent. In the last step a part
           of the polarized continuum is replaced by explicit solvent molecules.}
  \label{fig:solvationProcess}
\end{figure}

The resulting scheme is illustrated in Fig.~\ref{fig:solvationProcess}. As usual for PCM, the solvent and
solute are separated first, then a cavity in the continuum is created for the solute and the solute is inserted. This is
followed by the polarization of the continuum. Up to this point, the steps are identical to
the PCM. However, we now formally replace a part of the polarized continuum by
explicit solvent molecules. The interaction between these explicit solvent molecules and the
solute is now described with sDFT.
Hence, we have to consider the work spent on the polarization of the continuum in
the energy description of the sDFT solute--solvent interaction.
We chose such a substitution ansatz to avoid the calculation of the
entropy of the solvent molecules which is difficult to
capture with micro-solvation approaches. This is caused
by the fact that large spheres of solvent molecules can not be considered
as a supermolecule and their entropy is not recovered from approximations
such as the rigid-rotor/harmonic oscillator/particle in a box model commonly
used for single molecules\cite{Kamerlin2009}.

\subsection{Free Energy of Solvation in the Polarizable Continuum Model}

In the framework of the PCM the (Gibb's) free energy of solvation, \emph{i.e.}, the work
required for the process of bringing the solute $A$ from a separate phase into the
unperturbed solvent, may be expressed as\cite{tomasi2005quantum}
\begin{align}   
    \Delta G^A_\mathrm{solv} = E^A_\mathrm{solv} - E^A_0 + G_\mathrm{el} + G_\mathrm{cav} + G_\mathrm{dis-rep}+\Delta G_\mathrm{tm},
    \label{eq:freeEnergySolvPCM}
\end{align}
where $E^A_\mathrm{solv}$ and $E^A_0$ are the electronic total energies of $A$ in solution and
as an isolated species, respectively. $G_\mathrm{el}$, $G_\mathrm{cav}$, $G_\mathrm{dis-rep}$, and
$\Delta G_\mathrm{tm}$ are the free energy contributions due to electrostatic and polarization,
cavity formation, dispersion, and Pauli-repulsion, respectively, as well as change
in thermal motion of solute and solvent. The contributions $G_\mathrm{el}$,
$G_\mathrm{cav}$, and $G_\mathrm{dis-rep}$ are not denoted with a leading $\Delta$ since they
are not present in our reference.
For the solvent-related free energy contributions $G_\mathrm{el}$, $G_\mathrm{cav}$,
$G_\mathrm{dis-rep}$, expressions within the framework of PCM have been
formulated\cite{Mennucci2012}. For $\Delta G_\mathrm{tm}$ usually only the coordinates
of the solute are
considered\cite{tomasi2005quantum}, which may be evaluated
in the standard rigid rotator/harmonic oscillator/particle in a box approximation.
Note that we do not include a free energy contribution due to the change in
standard states between the initial gas phase and the final liquid phase since
we apply the so-called $\rho$-process from Ref.\citenum{ben1987solvation}.
This means that we formally separate the solute transfer from the gas phase
to solution into three steps.
First, we freeze the position of the solute in the gas phase. Then
we transfer it from its fixed position in the gas phase to a fixed position in the liquid phase, and finally,
we allow the solute to move in the liquid phase.
This separates the change of solute translation from the insertion into the solvent.
Furthermore, we assume that the solute translation is identical in
the gas phase and in solution, making the free energy contribution
from the translation change solely concentration dependent. 
For the $\rho$-process we select the concentrations in both phases to be identical. 
Hence, no additional free energy contribution arises.

Only $G_\mathrm{el}$ and
$G_\mathrm{dis-rep}$ in Eq.~(\ref{eq:freeEnergySolvPCM}) contain interaction energy
contributions between solute and solvent. Hence, we (partially) substitute the PCM
representation of these terms by the explicit sDFT-based model, for which we
first need to review the relevant parts of sDFT in the next section.

\subsection{Subsystem Density Functional Theory}
We consider a set of subsystems $\{I\}$ with $N$ elements, each of which is defined
by its density $\rho_I(\pmb{r})$ and a set of atoms with charges $\{Z_I\}$ at Cartesian coordinates
$\{\pmb{R}_I\}$. The total sDFT energy of such a system is given as
\begin{align}
  \begin{split}
    E_\mathrm{tot}^\mathrm{sDFT} =& \sum_I E^\mathrm{KS}_I[\rho_I(\pmb{r})]
     + \sum_{I\neq J} \int \mathrm{d}^3r~ v^I_\mathrm{nuc}(\pmb{r})\rho_J(\pmb{r})\\
     &+ \frac{1}{2} \sum_{I \neq J} \sum_{i\in I, j \in J} \frac{Z_i Z_j}{|\pmb{R}_i - \pmb{R}_j|}
     + \frac{1}{2} \sum_{I\neq J} \int \mathrm{d}^3r\mathrm{d}^3r^\prime~ \frac{\rho_I(\pmb{r})\rho_J(\pmb{r}^\prime)}{|\pmb{r}-\pmb{r}^\prime|}\\
     &+ T_s^\mathrm{nadd}[\{\rho_I(\pmb{r})\}] + E_\mathrm{xc}^\mathrm{nadd}[\{\rho_I(\pmb{r})\}],
  \end{split}
  \label{eq:sDFTEnergy}
\end{align}
where $v^I_\mathrm{nuc}(\pmb{r})$ is the Coulomb potential of the nuclei of system $I$ and $E^\mathrm{KS}_I[\rho_I(\pmb{r})]$
its KS-DFT energy. The notation $\sum_{I\neq J}$ implies
a summation over all $I$ and $J$, omitting the case $I = J$.
The functionals $T_s^\mathrm{nadd}[\{\rho_I(\pmb{r})\}]$ and
$E_\mathrm{xc}^\mathrm{nadd}[\{\rho_I(\pmb{r})\}]$ are the non-additive kinetic and the non-additive exchange--correlation
functional, respectively. They are defined by the non-interacting kinetic energy functional $T_s[\rho(\pmb{r})]$ and the
exchange--correlation functional $E_\mathrm{xc}[\rho(\pmb{r})]$ as
\begin{align}
  T_s^\mathrm{nadd}[\{\rho_I(\pmb{r})\}] = T_s^\mathrm{nadd}[\rho_1(r),\rho_2(r),...\rho_N(r)]
   = T_s\left[\sum_I \rho_I(\pmb{r})\right] - \sum_I T_s[\rho_I(\pmb{r})]
\end{align}
and
\begin{align}
  E_\mathrm{xc}^\mathrm{nadd}[\{\rho_I(\pmb{r})\}] = E_\mathrm{xc}^\mathrm{nadd}[\rho_1(\pmb{r}),\rho_2(\pmb{r}),...\rho_N(\pmb{r})]
   = E_\mathrm{xc}\left[\sum_I \rho_I(\pmb{r})\right] - \sum_I E_\mathrm{xc}[\rho_I(\pmb{r})].
\end{align}
We introduce a short hand notation for the intersubsystem electron--nuclei Coulomb interaction functional
$V_\mathrm{nuc,el}^{I,J}[\rho_I(\pmb{r}),\rho_J(\pmb{r})]$, the nuclei--nuclei Coulomb interaction energy
$V_\mathrm{nuc,nuc}^{I,J}$, and the electron--electron Coulomb interaction functional
$V_\mathrm{el,el}^{I,J}[\rho_I(\pmb{r}),\rho_J(\pmb{r})]$ for convenience as
\begin{align}
  V_\mathrm{nuc,el}^{I,J}[\rho_I(\pmb{r}),\rho_J(\pmb{r})] &= \int \mathrm{d}^3r~v^I_\mathrm{nuc}(\pmb{r})\rho_J(\pmb{r}) + \int \mathrm{d}^3r~v^J_\mathrm{nuc}(\pmb{r})\rho_I(\pmb{r}),\\
  V_\mathrm{nuc,nuc}^{I,J} &= \sum_{i\in I, j \in J} \frac{Z_i Z_j}{|\pmb{R}_i - \pmb{R}_j|},
\end{align}
and 
\begin{align}
  V_\mathrm{el,el}^{I,J}[\rho_I(\pmb{r}),\rho_J(\pmb{r})] = \int \mathrm{d}^3r\mathrm{d}^3r^\prime~\frac{\rho_I(\pmb{r})\rho_J(\pmb{r}^\prime)}{|\pmb{r}-\pmb{r}^\prime|},
\end{align}
and then for their sum as
\begin{align}
   V_\mathrm{el}^{I,J}[\rho_I(\pmb{r}),\rho_J(\pmb{r})] = V_\mathrm{nuc,el}^{I,J}[\rho_I(\pmb{r}),\rho_J(\pmb{r})]
+ V_\mathrm{nuc,nuc}^{I,J} + V_\mathrm{el,el}^{I,J}[\rho_I(\pmb{r}),\rho_J(\pmb{r})].
\end{align}

Since we are interested in the interaction between a specific subsystem $A$ (solute) and the union of all other
subsystems $B \equiv \bigcup_{I\neq A} I$ (solvent) we rewrite the energy decomposition accordingly as
\begin{align}
  \begin{split}
    E_\mathrm{tot}^\mathrm{sDFT} =& E^\mathrm{KS}_A[\rho_A(\pmb{r})] + E^\mathrm{KS}_B[\rho_B(\pmb{r})]
     +V_\mathrm{el}^{A,B}[\rho_A(\pmb{r}),\rho_B(\pmb{r})]\\
    &+ T_s^\mathrm{nadd}[\rho_A(\pmb{r}),\rho_B(\pmb{r})]
     + E_\mathrm{xc}^\mathrm{nadd}[\rho_A(\pmb{r}),\rho_B(\pmb{r})].
  \end{split}
\end{align}

Of course, this expression assumes a specific set of nucelar coordinates $\left\{ \{\pmb{R}_i\} \right\}^{(k)}$ in the Born--Oppenheimer
approximation. We will now indicate this set of coordinates by a configuration
index $k$.
The interaction energy between $A$ and $B$ is then defined as
\begin{align}
  E_\mathrm{int}^\mathrm{sDFT}(k) = V_{\mathrm{el},k}^{A,B}[\rho^k_A(\pmb{r}),\rho^k_B(\pmb{r})]
        +T_{s}^\mathrm{nadd}[\rho^k_A(\pmb{r}),\rho^k_B(\pmb{r})] + E_{\mathrm{xc}}^\mathrm{nadd}[\rho^k_A(\pmb{r}),\rho^k_B(\pmb{r})].
  \label{eq:sDFTInteractionEnergy}
\end{align}

\subsubsection{Electrostatic Interactions in the Hybrid Model}
In the PCM, the electrostatic free interaction energy between solute and solvent $G_\mathrm{el}$ is given by
\begin{align}
  \begin{split}
    G_\mathrm{el}^\mathrm{PCM} &=\frac{1}{2} V_\mathrm{el}^\mathrm{PCM}(\Gamma_{A})
                                      =\frac{1}{2}\int_{\Gamma_{A}} \mathrm{d}s~ \sigma(s) v_\mathrm{el}(s)
  \end{split}
  \label{eq:elecStatPCM}
\end{align}
where $\Gamma_{A}$ is the surface of the cavity created by the solute when inserted into the solvent,
$\sigma(s)$ is the (apparent) surface charge distribution on this surface and $v_\mathrm{el}(s)$ the electrostatic
potential of the embedded molecule\cite{Mennucci2012}. The free energy change corresponds to half
the electrostatic interaction $V_\mathrm{el}^\mathrm{PCM}(\Gamma_{A})$ between the polarized continuum and
the solute since our reference state (depicted in Fig.~\ref{fig:solvationProcess}) does not show any initial
polarization and we assume that the solvent ensemble average shows a linear response to turning-on electrostatic interactions which is reasonable for most solvents\cite{Aqvist1996}. Hence, the work needed for the polarization of the continuum corresponds to half of the solute--continuum electrostatic interaction\cite{tomasi1994molecular}.

An explicit representation of the polarized continuum by polarized solvent molecules allows us to
replace a part of the electrostatic solvent--solute interaction $V_\mathrm{el}^\mathrm{PCM}(\Gamma_{A})$
by its sDFT representation $\left\langle V_{\mathrm{el},k}^{A,B}[\rho^k_A(\pmb{r}),\rho^k_B(\pmb{r})]\right\rangle_k$ where $B$
is the union of all explicitly considered solvent molecules and $\langle ... \rangle_k$ denotes the averaging
over the configurations $k$. We then obtain for the combined electrostatic free energy
$G_\mathrm{el}$
\begin{align}
  \begin{split}
     G_\mathrm{el} &=\frac{1}{2} \left(V_\mathrm{el}^\mathrm{PCM}(\Gamma_{A+B}) + \left\langle V_{\mathrm{el},k}^{A,B}[\rho^k_A(\pmb{r}),\rho^k_B(\pmb{r})]\right\rangle_k\right).
  \end{split}
  \label{eq:elecStatPCMandsDFT}
\end{align}
However, since the explicit expression of $V_\mathrm{el}^\mathrm{PCM}$ [Eq.~(\ref{eq:elecStatPCM})]
depends on the now enlarged cavity surface $\Gamma_{A+B}$, $V_\mathrm{el}^\mathrm{PCM}(\Gamma_{A+B})$ depends on
the configuration $k$, \emph{i.e.},
\begin{align}
   V_\mathrm{el}^\mathrm{PCM}(\Gamma_{A+B}) \rightarrow V_\mathrm{el}^\mathrm{PCM}(\Gamma_{A+B};k).
\end{align}
Therefore, it is necessary to add $V_\mathrm{el}^\mathrm{PCM}(\Gamma_{A+B};k)$ before the averaging
over $k$ as
\begin{align}
  \begin{split}
  \end{split}
     G_\mathrm{el} =\frac{1}{2} \left\langle V_\mathrm{el}^\mathrm{PCM}(\Gamma_{A+B};k) + V_{\mathrm{el},k}^{A,B}[\rho^k_A(\pmb{r}),\rho^k_B(\pmb{r})]\right\rangle_k .
  \label{eq:elecStatPCMandsDFT_averaged}
\end{align}

\subsection{The Electronic Energy Functional}
Deriving the energy functional for the non-additive exchange--correlation and kinetic energy
contributions is more complicated. Therefore, we resort to a more general approach by analyzing the
effective solute Hamiltonian $\hat{H}^\mathrm{eff}[\Phi]$ of our sDFT-hybrid model associated to
the solute wavefunction $\Phi$ as done for the PCM\cite{tomasi2005quantum}.
We will do this for a given solvent configuration $k$ and then
average over all configurations to derive the final energy expression.
The effective solute Hamiltonian is given by
the Hamiltonian of the isolated solute $\hat{H}_0$ and the interaction operators for the solvent--solute interaction
\begin{align}
  \hat{H}^\mathrm{eff}[\Phi] = \hat{H}_0 + \hat{V}_\mathrm{el}^S[\Phi] + \hat{V}_\mathrm{xc}^S[\Phi] + \hat{V}_\mathrm{kin}^S[\Phi],
\end{align}
where we have denoted the electrostatic Coulomb interaction operator with the solvent as $\hat{V}_\mathrm{el}^S[\Phi]$,
the exchange--correlation operator as $\hat{V}_\mathrm{xc}^S[\Phi]$, and the non-additive kinetic energy operator
as $\hat{V}_\mathrm{kin}^S[\Phi]$.
We now aim at solving  a
Schrödinger equation with eigenvalue $\epsilon$ of the form
\begin{align}
  \hat{H}^\mathrm{eff}[\Phi]~\Phi = \epsilon~\Phi,
  \label{eq:nonLinearSchroedinger}
\end{align}
and construct the energy functional $E[\Phi]$ whose stationary points correspond to the solutions or approximations to the solution of the functions $\Phi$.
Eq.~(\ref{eq:nonLinearSchroedinger}) is non-linear in $\Phi$ because the interaction operators depend
on this solution function. A general discussion for such non-linear Schrödinger equations is given
in Ref.~\citenum{Sanhueza1979}.
We will express the operators $\hat{V}_\mathrm{el}^S[\Phi]$, $\hat{V}_\mathrm{xc}^S[\Phi]$, and $\hat{V}_\mathrm{kin}^S[\Phi]$ in the form $\hat{V}^{(q)}[\Phi] = \hat{A}(\Phi^*
\Phi)^q$, where $\hat{A}$ is an operator independent of $\Phi$.

Hence, the energy functional $E[\Phi]$ may be written as a sum
over energy contributions for each operator. These energy
contributions $E^{(q)}[\Phi]$ depend on the exponent $q$ that enters the operator, since it determines the occurrence of $\Phi$ in the operator which is varied when solving
Eq.~(\ref{eq:nonLinearSchroedinger}). This energy contribution is given as
\begin{align}
  E^{(q)}[\Phi] = \frac{\left\langle \Phi \left|\hat{V}^{(q)}[\Phi] - \frac{q}{q+1} \hat{V}^{(q)}[\Phi] \right| \Phi \right\rangle}{\langle \Phi | \Phi \rangle}.
\end{align}
The scaling of the operator by the factor $1-\frac{q}{q+1}$ corrects the additional terms which arise for a
variation $\Phi \rightarrow \Phi + \delta \Phi$ due to the occurrence of $\Phi$ in $\hat{V}^{(q)}[\Phi]$
\cite{Sanhueza1979}.

As stated in the previous section, the electrostatic interaction is caused by the solvent polarization which is modeled to be linear in
$\Phi^* \Phi$ in the PCM. Hence, the factor of $1/2$ emerges in the expression of the electrostatic free energy contribution [see Eq.~(\ref{eq:elecStatPCMandsDFT_averaged})]. For the exchange--correlation
operator $\hat{V}_\mathrm{xc}^S[\Phi]$ and the non-additive kinetic energy operator we assume that they have the form of a
local potential which can be calculated from sDFT for the non-additive exchange--correlation and kinetic energy as
\begin{align}
  \hat{V}_\mathrm{xc/kin}^S[\Phi] &= \sum_i \left\{v_\mathrm{xc/kin}[\rho_A(\pmb{r}) + \rho_B(\pmb{r})](\pmb{r}_i) - v_\mathrm{xc/kin}[\rho_A(\pmb{r})](\pmb{r}_i)\right\},
  \label{eq:localPotentialOperator}
\end{align}
where $\pmb{r}_i$ is the Cartesian coordinate of electron $i$ and $v_\mathrm{xc/kin}[\rho(\pmb{r}^\prime)](\pmb{r})$ is either the exchange--correlation ($\mathrm{xc}$)
or the kinetic energy ($\mathrm{kin}$) potential.

We now consider the exchange part $v_\mathrm{x}[\rho](\pmb{r}_i)$ of the local density
approximation (LDA) for the exchange--correlation potential. In LDA, the first term in Eq.~(\ref{eq:localPotentialOperator})
is given by\cite{YangParr1994}
\begin{align}
  v_\mathrm{x}[\rho_A+\rho_B](\pmb{r}_i) = \frac{4}{3} C_x \left[\rho_A(\pmb{r}_i)+\rho_B(\pmb{r}_i)\right]^{1/3}
  \label{eq:firstTermNonAdditive}
\end{align}
and the second term by
\begin{align}
  v_\mathrm{x}[\rho_A](\pmb{r}_i) = \frac{4}{3} C_x \rho^{1/3}_A(\pmb{r}_i),
\end{align}
where $C_x = 3/4 \left( 3/\pi \right)^{1/3}$. Since we have $\rho_A(\pmb{r}) = \left\langle \Phi\left|\sum_i \delta(\pmb{r}_i-\pmb{r})\right| \Phi\right\rangle = \hat{A}(\Phi^{*} \Phi)^1$,
we can identify $q=1/3$ for the second term. For the first term, however, such a relation is not immediately clear, because of $\rho_B(\pmb{r}_i)$ in the operator.
For the electrostatic interaction the total charge distribution of $B$ is linear in $\Phi^*\Phi$
($q=1$). For simplicity, we assume that also the electron density is a linear function of $\Phi^*\Phi$
\begin{align}
    \rho_B(\pmb{r}_i) \approx \hat{O}\Phi^*\Phi,
    \label{eq:rhoBLinearInPhi}
\end{align}
where $\hat{O}$ is a not further specified operator independent of $\Phi$.
Note that this is an approximation since the non-polarized continuum shows a homogeneous, non-zero electron density distribution in
the absence of the solute. However, the electron density of the explicitly considered solvent molecules is far from homogeneous, which means
that the polarization by the solute dominates the electron density distribution and we may assume that
$\rho_B(\pmb{r}_i) \approx \hat{O}\Phi^*\Phi$ is sufficiently accurate at least in the spatial region where $\Phi$ is large. Furthermore, we know from the electrostatic interaction that there
must be some dependence on $\Phi^*\Phi$ which we have to include. We assume $q=1/3$ for Eq.~(\ref{eq:firstTermNonAdditive}) and hence for
$\hat{V}_\mathrm{xc}^S[\Phi]$.
We will not derive an explicit expression for the correlation part of the operator since we will be unable to separate exchange and correlation in
the final calculations.

For many exchange--correlation functional approximations in KS-DFT and sDFT the semi-empirical D3 dispersion
correction\cite{Grimme2010} is parameterized and may be used as an energy correction on top of the density dependent
functional. The D3 correction takes the form of atom-dependent two- and three-body terms. As such, it has no
direct representation in the effective Hamiltonian. However, we can still analyze its leading energy
contribution in terms of its dependence on $\Phi^*\Phi$ and compare it to the expression for the electrostatic
interaction. The leading energy contribution in the D3 correction is given as the two-body term between two atoms $A$ and
$B$ with distance $r_{AB}$, dispersion coefficient $C^{AB}_n$ and damping function $f_{d,6}(r_{AB})$ as
\begin{align}
    E^{(2)}_{D3} = \frac{C_6^{AB}}{r_{AB}^6} f_{d,6}(r_{AB})~.
\end{align}
The damping function $f_{d,6}(r_{AB})$ was introduced to avoid divergence of $E^{(2)}_{D3}$ for small $r_{AB}$.
Hence, we assume $f_{d,6}(r_{AB})$ to be independent of $\Phi^*\Phi$. This leaves only the $C_6^{AB}$-coefficient to be analyzed.
It may be written in terms of the frequency dependent polarizablities $\alpha^A(i\omega)$ of $A$ and
$\alpha^B(i\omega)$ of $B$ as\cite{Casimir1948}
\begin{align}
    C_6^{AB} = \frac{3}{\pi} \int_0^\infty \alpha^A(i\omega) \alpha^B(i\omega)\mathrm{d}\omega~.
\end{align}
Note that the polarizabilities $\alpha^{A/B}(i\omega)$ are linear in the wavefunction products
$\Phi_{A/B}^*\Phi_{A/B}$. If we now consider our solute instead of an atom $A$ and the solvent instead of an atom $B$ and, in analogy to Eq,~(\ref{eq:rhoBLinearInPhi}), model $\Phi_{B}^*\Phi_{B}$ to be
a linear function of $\Phi^*\Phi$, we find that $E^{(2)}_{D3}$ is quadratic in $\Phi^*\Phi$. The same quadratic dependency is given for the electrostatic
interaction which is multiplied by $1/2$ in its solvation free energy contribution [see Eq.~\ref{eq:elecStatPCMandsDFT_averaged}] to account for the solvent polarization. Hence, we must introduce the same factor of $1/2$ for the D3-based dispersion
correction.

For the non-additive kinetic energy operator, we construct a model
similar to that of the density-based exchange--correlation interaction
operator with the Thomas--Fermi model\cite{YangParr1994} for the
kinetic energy potential, which yields an exponent of $q = 2/3$.

The total energy is then given by
\begin{align}
    E[\Phi] &= {\left\langle \Phi \left| \hat{H}_0 \right| \Phi \right\rangle}
             + E_\mathrm{el}^A + E_\mathrm{xc}^A + E_\mathrm{Disp}^A + E_\mathrm{kin}^A,
\end{align}
with interaction energy contributions defined as
\begin{align}
\begin{split}
    E_\mathrm{el}^A &= \frac{1}{2} {\left\langle \Phi \left| \hat{V}_\mathrm{el}^S[\Phi] \right| \Phi \right\rangle},\\
    E_\mathrm{xc}^A &= \frac{3}{4} {\left\langle \Phi \left| \hat{V}_\mathrm{xc}^S[\Phi] \right| \Phi \right\rangle},\\
    E_\mathrm{disp}^A &= \frac{1}{2} \left(E^{D3}_{A+B} - E^{D3}_A - E^{D3}_B \right),\\
    E_\mathrm{kin}^A &= \frac{3}{5} {\left\langle \Phi \left| \hat{V}_\mathrm{kin}^S[\Phi] \right| \Phi \right\rangle},
\end{split}
  \label{eq:finalEnergyScaling}
\end{align}
for normalized $\Phi$, \emph{i.e.}, $\left\langle \Phi|\Phi\right\rangle = 1$.
The operators in these expectation values are
\begin{align}
  \begin{split}
    \hat{V}_\mathrm{el}^S[\Phi] &= -\sum_i \int \mathrm{d}^3r^\prime \frac{p_B[\Phi](\pmb{r}^\prime)}{|\pmb{r}_i-\pmb{r}^\prime|}
    							 + \sum_I Z_I \int \mathrm{d}^3r^\prime \frac{p_B[\Phi](\pmb{r}^\prime)}{|\pmb{R}_I-\pmb{r}^\prime|},\\
    \hat{V}_\mathrm{xc}^S[\Phi] &= \sum_i \left\{v_\mathrm{xc}[\rho_A(\pmb{r}) + \rho_B(\pmb{r})](r_i) - v_\mathrm{xc}[\rho_A(\pmb{r})](r_i)\right\},\\
    \hat{V}_\mathrm{kin}^S[\Phi]&= \sum_i \left\{v_\mathrm{kin}[\rho_A(\pmb{r}) + \rho_B(\pmb{r})](r_i) - v_\mathrm{kin}[\rho_A(\pmb{r})](r_i)\right\}.
  \end{split}
\end{align}
$E^{D3}_{A+B}$, $E^{D3}_{A}$, and $E^{D3}_{B}$ are D3 corrections calculated with the union of all solvent and
solute atoms ($A+B$) and calculated only for solute ($A$) and solvent ($B$), respectively.
The charge distribution, including the sign information of the charges, created by the PCM charges and the explicitly considered solvent molecules is denoted by
$p_B[\Phi]$.

In the following we will introduce the configuration index $k$ for the electrostatic interaction and rewrite the dependence
of the energy functionals as functionals of the electron density of the solute and the density of the explicitly
considered solvent molecules. Furthermore, we introduce expressions for the exchange--correlation  interaction ($E^{\mathrm{PCM},k}_\mathrm{xc}$, $E^{\mathrm{PCM},k}_\mathrm{disp}$) and non-additive kinetic energy ($E^{\mathrm{PCM},k}_\mathrm{kin}$)
with the remaining continuum that was not substituted by explicit solvent molecules, since we do not have any direct representation of its electron density
\begin{align}
  \begin{split}
    E_\mathrm{el}^A &\rightarrow E_\mathrm{el}^{A,k}[\rho_A; \rho_B]\\
    E_\mathrm{xc}^A &\rightarrow E_\mathrm{xc}^{A,k}[\rho_A^k;\rho_B^k] + E^{\mathrm{PCM},k}_\mathrm{xc}\\
    E_\mathrm{disp}^A &\rightarrow E_\mathrm{disp}^{A,k} + E^{\mathrm{PCM},k}_\mathrm{disp}\\
    E_\mathrm{kin}^A &\rightarrow E_\mathrm{kin}^{A,k}[\rho_A^k;\rho_B^k] + E^{\mathrm{PCM},k}_\mathrm{kin}.
  \end{split}
\end{align}
The terms $E_\mathrm{xc}^{A,k}[\rho_A^k;\rho_B^k]$, $E_\mathrm{disp}^{A,k}$ and
$E_\mathrm{kin}^{A,k}[\rho_A^k;\rho_B^k]$ can be evaluated with the wavefunction $\Phi$
that corresponds to $\rho_A^k$ as shown in
Eq.~(\ref{eq:finalEnergyScaling}) while expressions known for the PCM\cite{Floris1991,Floris1993,Amovilli1997,Weijo2010} may be used for
$E^{\mathrm{PCM},k}_\mathrm{xc}$, $E^{\mathrm{PCM},k}_\mathrm{disp}$,
and $E^{\mathrm{PCM},k}_\mathrm{kin}$.

\subsection{Solvation Free Energy of the Hybrid Model}

With the definitions above, the final result for the free energy of solvation
of our sDFT-hybrid model is given by
\begin{align}
  \begin{split}
  \Delta G_\mathrm{solv}^{A} = &\left\langle E^\mathrm{KS}_A[\rho^k_A] 
     + E_\mathrm{kin}^A[\rho_A^k;\rho_B^k] + E_\mathrm{xc}^A[\rho_A^k;\rho_B^k] + E_\mathrm{disp}^{A,k}
     + E_\mathrm{el}^{A,k}[\rho_A^k;\rho_B^k]\right. \\&+\left. 
     E^{\mathrm{PCM},k}_\mathrm{el}[\rho_A^k;\rho_B^k]
     \right\rangle_k
      - E^\mathrm{KS}_A[\rho^0_A]
          + \Delta G_\mathrm{tm}^\mathrm{A} + \Delta G_\mathrm{cav}\\
          &+ \Delta G_\mathrm{tm}^\mathrm{S} + \langle E^{\mathrm{PCM},k}_\mathrm{xc} + E^{\mathrm{PCM},k}_\mathrm{kin} + E^{\mathrm{PCM},k}_\mathrm{disp}\rangle_k~.
  \end{split}
  \label{eq:finalSolvationFreeEnergy}
\end{align}
The term $\Delta G_\mathrm{tm}^\mathrm{A}$ [see Eq.~(\ref{eq:freeEnergySolvPCM})] denotes the free energy contributions from the change in thermal
motion of the solute and the term $\Delta G_\mathrm{cav}$ denotes the change in free energy
due to cavity creation. For the latter, expressions from the PCM can be employed\cite{Mennucci2012}.

The energy definition in Eq.~(\ref{eq:finalSolvationFreeEnergy}) does not require an
energy calculation for the solvent while still including the polarization work done in the environment.
Furthermore, it only requires the electron density of the explicitly considered molecules in spatial regions
close to the solute to evaluate the exchange--correlation interaction and non-additive kinetic energy interaction.
Only in these regions is the solute electron density significant. Therefore, these terms will converge quickly with the number of explicitly
considered solvent molecules.

We note here that
this final expression is similar to the expression obtained in the so-called ``shells theory
of solvation''\cite{Pliego2010}, especially in its form in Refs.~\citenum{Lima2010} and \citenum{Lima2011}.
In the ``shells theory of solvation'' approach, the response of the solvent
ensemble average is assumed to be linear for the total interaction
between solvent and solute. This leads to a factor of $1/2$ for the interaction
energy. In this work, however, we assume a more complicated relationship between solute and solvent and employed
non-linear terms in the exchange--correlation and non-additive kinetic energy
functionals.

\subsection{Sampling Configurations of the Explicit Solvent}
The addition of explicit solvent molecules described in full atomic resolution creates the burden of sampling these orientations.
We probe two different approaches towards sampling.
On the one hand, we employ molecular dynamics (MD) simulations to generate time distributed snapshots.
On the other hand, we randomly generate and then optimize static snapshots.
The details of both procedures are given in the next section.

First, we discuss the differences, benefits, and short-comings of both approaches.
One key difference is the way the individual snapshots are combined to model the partition function of the total system.
The MD approach includes thermodynamic integration, meaning that equidistant (in time) sampling should ultimately represent an accurate sample of the exact system.
The snapshots extracted are equally weighted, and prevalence of a specific configuration is represented by the same or similar snapshot appearing multiple times in the sample.
In contrast, the static approach should not yield the same configuration twice and instead requires an accurate, non-uniform weighting of the different configurations afterwards.
The weighting follows a Boltzmann distribution for the probability $w_k$ of finding a configuration $k$ given as
\begin{align}
    w_k=\frac{1}{Z}\text{exp}\left\{\frac{-(E_{k}-E_\text{min})}{k_\text{B}T}\right\}.
    \label{eq:boltzmannWeight}
\end{align}
Here, $T$ is the temperature, $k_\text{B}$ the Boltzmann constant, $E_k$ is the energy derived from the solute energy $E^\mathrm{KS}_A[\rho^k_A]$ of the configuration $k$, $E_\text{min}$ is the minimal energy of all $E_k$s and $Z$ the partition function, \begin{align}
    Z=\sum_{k}\text{exp}\left\{\frac{-(E_k-E_\text{min})}{k_\text{B}T}\right\}.
    \label{eq:partitionFunction}
\end{align}
The energy $E_k$ is given by
\begin{align}
    E_k = E^\mathrm{KS}_A[\rho^k_A] + E^{\mathrm{PCM},k}_\mathrm{el}[\rho^k_A]
    \label{eq:modSoluteEnergy}
\end{align}
with $\rho^k_A$ corresponding to the density of the solute in a solute--solvent cluster and $E^\text{PCM,k}_\mathrm{el}[\rho^k_A]$ being the electrostatic interaction energy of the solute with density $\rho^k_A$ with the PCM.
This definition of $E_k$ focuses on the solute while accounting for the perturbation by the solvent, as attempted previously.\cite{Simm2020} 
We refer to this sampling as weighted random minimum structures (WRMS).
With WRMS, observables of the total system can be obtained for the optimized static snapshots.\\

Both approaches require to be repeated if either solvent composition or solute change, but they differ in this point when the temperature is changed.
While the MD simulation yields a different trajectory at different temperature, the static WRMS picture can simply be re-weighted with the new temperature.
The employed optimization, if performed rigorously, will yield only local minima.
The assumption that WRMS describe the exact state function accurately may be reasonable for low temperatures, however, for higher temperatures the state function will more and more become a continuum that is not accurately modeled by this approach.
On top of these conceptual differences between the two approaches, they also differ in the models employed to describe the atomic systems.
The WRMS approach uses a micro droplet to model the water at any time, while the MD simulation was run using periodic boundary conditions.
Foreshadowing the results a little, it can be expected that this will be one factor that leads to different behaviour when determining the amount of explicit solvent that has to be modeled in order to obtain converged results.
Additionally, the differences in methods chosen for the energy and force evaluation is also of interest.
In the following consideration, we shall require that the generation of the samples should be relatively fast.
For this reason, we will not discuss DFT or sDFT structure optimizations or ab-initio molecular dynamics. 
The remaining methods that are commonly available are force fields and semi-empirical methods, such as tight-binding (\emph{e.g}. GFN2-xTB\cite{Bannwarth2019}) and NDDO type methods (\emph{e.g} PM6\cite{Stewart2007}).
The usage of standard force fields will be the computationally quickest solution, but will require that both solvent and solute are modeled by the force field, or that the required parameters can readily be generated.\cite{Grimme2014,Brunken2020}
Semi-empirical quantum chemistry methods are of key interest then.
However, their application to several nanoseconds of MD simulations is already beyond the computational effort we deem 'relatively fast'.
Assuming that a single trajectory of several nanoseconds is required, the time needed to generate this trajectory is far greater than the time required to evaluate the snapshots,
the main reason being that the snapshot evaluation is easily parallelizable without a computational diminishing return.

\section{Computational Methodology}

\subsection{Example Reactions}
To investigate our hybrid solvation model, we consider the Menshutkin reaction of ammonia with chloromethane in water $\left( \ce{NH3 + CH3Cl \rightarrow [NH3CH3]+ + Cl-}\right)$
which is frequently studied to asses the reliability of solvation models,
ranging from FEP\cite{Gao1991,Chen2019}, effective fragment
potentials,\cite{Webb1999} implicit solvation models based on COSMO\cite{Truong1997},
PCM\cite{Amovilli1998,Castejon1999,Su2007,Chuang1998}, and the reference interaction side model\cite{Naka1999}.
Menshutkin-type reactions are S$_\text{N}$2-type reactions where the reactants are neutral while leading to a significant change
in partial charges along the reaction coordinates. 
The transition state already shows significant ionic character which produces a pronounced solvent effect on the reaction barrier. 
Since no explicit experimental data is available for the reaction of ammonia with chloromethane, the experimental reaction
barrier of its iodine variant $\left( \ce{NH3 + CH3I \rightarrow [NH3CH3]+ + I-}\right)$\cite{Okamoto1967} is considered
as a lower estimate since iodine is a better leaving group than chloride.
Notably, it was found in Ref.~\citenum{Chen2019} that the FEP-based approaches using the CHARMM generalized force
field\cite{Vanommeslaeghe2009} fail by underestimating this lower estimate and must be corrected with accurate
quantum chemical methods in a second step.

\begin{figure}[ht]
    \centering
    \includegraphics{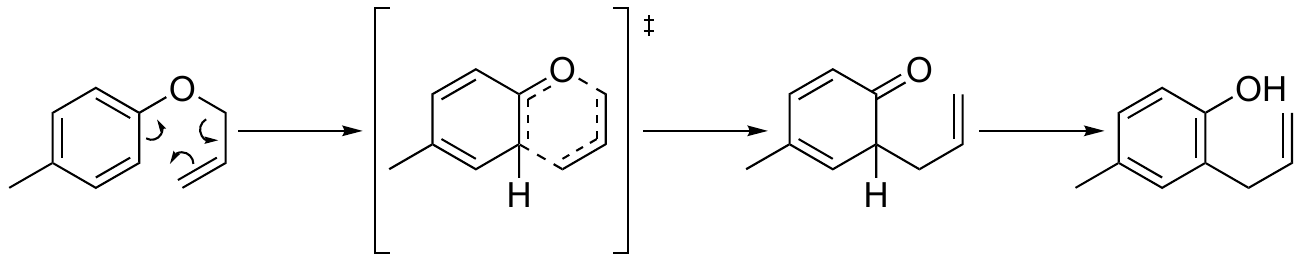}
    \caption{Claisen rearrangement of p-tolyl ether.}
    \label{fig:lewisClaisen}
\end{figure}

As a second example, we consider the Claisen rearrangement of p-tolyl ether in water and in cyclohexane
as shown in Fig.~\ref{fig:lewisClaisen}. 
The first step of the reaction, the actual rearrangement, is slow and therefore the rate determining step. 
The tautomerization in the second step of the reaction is fast. 
Therefore, we will consider the first step only. 
Experimentally, the kinetics of this reaction have been studied, amongst other solvents, in an ethanol-water mixture and in tetradecane.\cite{White1970}
For this work, we chose to simplify the binary mixture by pure water, the main component of the mixture, and tetradecane by cyclohexane, as they have very similar dielectric constants\cite{Wohlfarth2008,Wohlfarth2008a} and the choice of latter reduces the computational cost. 
This reaction was also studied in previous work with FEP\cite{Acevedo2010} and implicit solvation\cite{irani2009joint}.
Interestingly, in Ref.~\citenum{irani2009joint} the PCM was reported to be unable to predict the change in the reaction barrier between gas
phase, non-polar and polar solvents as indicated by the change in reaction rate measured in the experiment\cite{White1970}.

\subsection{Molecular Structure Optimizations\label{sec:MolStrucOpt}}
The transition state and the isolated reactants were optimized with
the program package \textsc{Turbomole}\cite{TURBOMOLE}, applying the PBE0 exchange--correlation functional\cite{Adamo1999} in a def2-TZVP basis set\cite{Ahlrich2005},
Grimme's D3 dispersion correction\cite{Grimme2010} with Becke--Johnson
damping\cite{Grimme2011} and the COSMO implicit solvation model\cite{Klamt2017}. The
dielectric constant was chosen to be $78.39$ for water and $2.00$ for cyclohexane.

\subsection{Molecular Dynamics based Micro-Solvation Setup}
The MD simulations for the solvent geometry generation
were carried out as follows:
Langevin dynamics simulations were employed with periodic boundary conditions and
the CHARMM General Force Field (CGenFF)\cite{Vanommeslaeghe2009,Yu2012} and
atom types were assigned with the cgenff.paramchem.org online-tool\cite{Vanommeslaeghe2012, Vanommeslaeghe2012a}. In the case of water
as solvent, the TIP3P water model\cite{Jorgensen1983} was chosen. 
For the Menshutkin reaction, the simulation box had a volume of $36~\si{\angstrom}
\times 26~\si{\angstrom} \times 26~\si{\angstrom}$. For the Claisen rearrangement
of p-tolyl-ether, the box was chosen as $30~\si{\angstrom}
\times 30~\si{\angstrom} \times 30~\si{\angstrom}$.
The structures of Section~\ref{sec:MolStrucOpt} were placed in the center
of each box. The box was then filled with water molecules
using the program VMD\cite{Humphrey1996} ($720$ molecules for the Menshutkin reaction and $804$ molecules for the Claisen rearrangement) and in case of cyclohexane as the solvent with
the program packmol\cite{Martinez2009} (150 molecules in the $30~\si{\angstrom}
\times 30~\si{\angstrom} \times 30~\si{\angstrom}$ box).
The atomic charges for the structures were replaced by the Merz--Kollmann electrostatic
potential charges\cite{Singh1984} calculated with \textsc{Turbomole} and PBE0/def2-TZVP/COSMO as chosen for the structure optimization.
The program \textsc{NAMD}\cite{Phillips2005} was applied for the MD simulations for an NPT ensemble ($1~\si{atm}, 25\si{\degree C}$) for $6~\si{ns}$ each,
keeping the coordinates of the
solute molecules frozen. A time step of $2~\si{fs}$ was chosen and the bonds of the
solvent molecules were kept rigid.

From the last $4~\si{ns}$ of the MD trajectories, configurations were extracted at random. Molecular
clusters were constructed by retaining only the closest solvent molecules up to a specified
number. Then, the solvation free energy was calculated in a sDFT calculation.
For each solute molecule the sampling occurred in batches of $32$ frames until the
change in solvation free energy between two continuous batches was below $0.1~\si{kJ.mol^{-1}}$.

\subsection{Optimization based Micro-Solvation Setup}
\begin{figure}[ht]
    \centering
    \includegraphics[width = 0.5\textwidth]{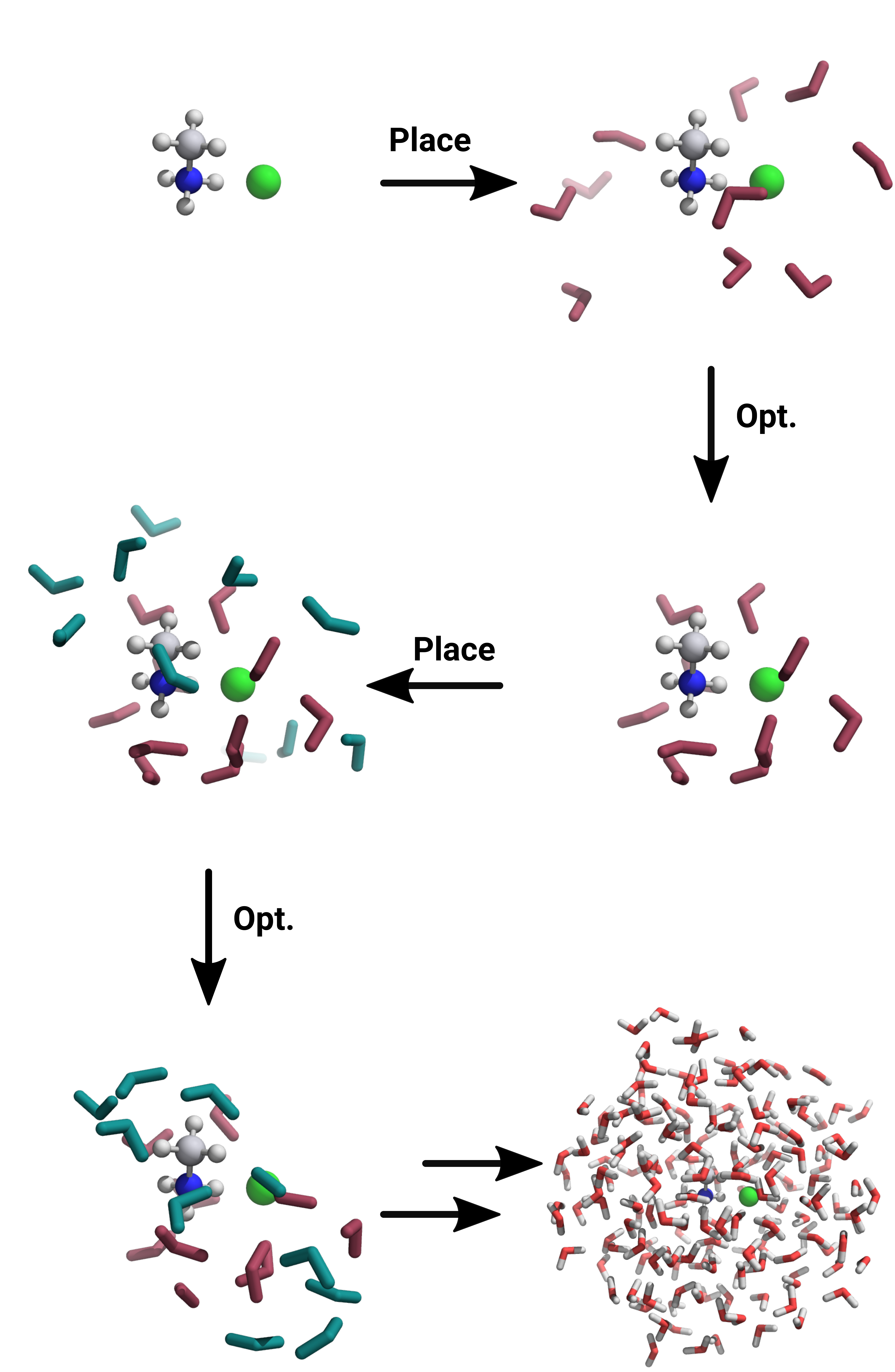}
    \caption{Solute--solvent clusters after the placement of solvent molecules to the product of the Menshutkin reaction and clusters after the optimization based on GFN2-xTB. The prodcut of the Menshutkin reaction is represented as ball-and-stick model, the first ten water molecules are represented in red, the second ten water molecules are represented in blue. The final solute--solvent cluster contains 180 water molecules.}
    \label{fig:build_up}
\end{figure}

All initial solvent clusters were generated by a modified version of the algorithm described by Simm \emph{et al}\cite{Simm2020}.
For a cluster with, \emph{e.g.}, 10 solvent molecules, the solvent molecules are placed around the solute, as can be seen in Fig.~\ref{fig:build_up}, based on the classification of the surface points of the solute in two categories: open and covered sites.
A ray originating from an open site does not hit any solvent molecule, a ray originating from a covered site does.
Solvent molecules are added until the requested number is reached.
During the placement procedure, the solute is redefined, meaning that all present solvent molecules are considered being part of the solute, if more than \SI{85}{\%} of the initial surface points are considered covered.
This guarantees a spherical distribution of solvents around the original solute.
The solute--solvent cluster is then optimized, as shown in Fig.~\ref{fig:build_up}, while keeping the coordinates of the original solute molecule frozen.
For the next cluster of the series with, for instance, a total of 20 solvent molecules, solvent molecules are added to the optimized cluster according to the procedure described above and the optimization is repeated.
This is continued in an automatized fashion until the total number of solvent molecules is reached.
Here, one could implement a check for convergence of $\Delta G_\textrm{solvation}$ after each addition of solvent molecules and terminate, if the targeted convergence criteria, for instance a deviation of less than \SI{4.18}{kJ/mol} of the observed energy over the last three points, are reached.

The final source code used to place the solvent molecules is released with the current version of the \textsc{SCINE Utilities}\cite{ScineUtils}. 
For GFN-FF\cite{Spicher2020} and also GFN2-xTB\cite{Bannwarth2019}, their electronic energies and nuclear gradients were forwarded to \textsc{SCINE Readuct}\cite{Readuct, ScineReaduct} for the optimization.
The two methods are available in \textsc{SCINE Readuct} through a wrapper around \textsc{xTB}\cite{Bannwarth2020}.
For CGenFF, the clusters were optimized with the optimize functionality of the \textsc{NAMD} program\cite{Phillips2005} employing CGenFF with the same parameters as described for the MD.
For each solute--solvent cluster, 224 configurations served as starting points for each cluster series. 

\subsection{Subsystem Density Functional Theory-based Solvation Free Energies}
All subsystem density functional theory calculations were performed with a development version of \textsc{Serenity} based on version
v1.3.1\cite{Unsleber2018, SerenityZenodo}.
The calculation of the energy expression $E^{\mathrm{PCM},k}_\mathrm{el}[\rho^k_A]$ needed for the
Boltzmann weighting of the optimized solvent clusters (see Eq.~\ref{eq:modSoluteEnergy}) was added to the already released code.
A patch for this code can be requested, and will be part of the next release of the \textsc{Serenity} program.

For the sake of consistency with computational methods applied for the structure
optimization, the PBE0 exchange--correlation functional\cite{Adamo1999} was employed
for calculations on the solute, whereas the PBE\cite{perdewBurkeErnzerhof}
functional in KS-DFT calculations for solvent molecules and as the non-additive
exchange--correlation functional. For both exchange--correlation functionals
the D3 dispersion correction\cite{Grimme2010} with Becke--Johnson 
damping\cite{Grimme2011} was employed. Lembarki and Chermette's kinetic energy
functional known as PW91k\cite{lembarki94_pw91k} delivered the non-additive
kinetic energy. The solute--solvent clusters were embedded in CPCM\cite{Barone1998} (i) with a static permittivity of $78.39$ and a solvent probe
radius of $2.6$ Bohr to describe water and (ii) with a static permittivity of $2.0$ with
a probe radius of $5.3$ Bohr to describe cyclohexane.
The molecular cavities were constructed with the scheme proposed by
Delley\cite{Delley2006} with atomic radii by
Bondi\cite{bondi1964van} scaled by a factor of $1.2$. The cavity formation
energy $\Delta G_\mathrm{cav}$ [see Eq.~(\ref{eq:finalSolvationFreeEnergy})] was calculated as according to Ref.~\citenum{langlet1988improvements}.
To mimic the solvent effect on the harmonic frequencies of the
solute (contributes to $\Delta G_\mathrm{tm}^A$), frequency calculations were carried out for the solute embedded in CPCM.
Hessians were obtained semi-numerically with \textsc{SCINE Readuct} from aggregating gradients calculated with \textsc{Serenity}.
In the following, $\Delta G_\mathrm{solv}^{A}$ defined in Eq.~(\ref{eq:finalSolvationFreeEnergy}) is approximated as

\begin{align}
  \begin{split}
  \Delta G_\mathrm{solv}^{A} \approx \Delta G_\mathrm{solvation} = &\left\langle E^\mathrm{KS}_A[\rho^k_A] 
     + E_\mathrm{kin}^A[\rho_A^k;\rho_B^k] + E_\mathrm{xc}^A[\rho_A^k;\rho_B^k] \right.\\
     & + \left. E_\mathrm{disp}^{A,k} +E_\mathrm{el}^{A,k}[\rho_A^k;\rho_B^k] + 
     E^{\mathrm{PCM},k}_\mathrm{el}[\rho_A^k;\rho_B^k]
     \right\rangle_k\\
     & + \Delta G_\mathrm{tm}^\mathrm{A} + \Delta G_\mathrm{cav} - E^\mathrm{KS}_A[\rho^0_A].\\
 \end{split}
  \label{eq:finalApproxSolvationFreeEnergy}
\end{align}

The approximations include not considering the change in thermal motion of the solvent, $\Delta G_\text{tm}^\text{S}$, and the terms $E^{\mathrm{PCM},k}_\mathrm{xc}$, $E^{\mathrm{PCM},k}_\mathrm{disp}$,
and $E^{\mathrm{PCM},k}_\mathrm{kin}$.

\section{Results and Discussion}

Weighted random minium structures were derived with GFN-FF, GFN2-xTB and CGenFF, whereas structures of the MD were only obtained from CGenFF.
The performance of these four combinations of sampling approaches and electronic structure methods is now examined with the Menshutkin reaction in water and the Claisen rearrangement in water and cyclohexane.

\subsection{Convergence with the Number of Explicit Solvent Molecules}

To assess internal consistency of our hybrid solvation model, we first need to analyze the convergence of the solvation free energies in the corresponding solvent of all species involved in all of our example reactions.
This includes the reagents, the transition states, and the products.
We start with analyzing $\Delta G_\text{solvation}$ of these in the Menshutkin reaction.

\begin{figure}[!ht]
    \centering
    \includegraphics[width = 0.975\textwidth]{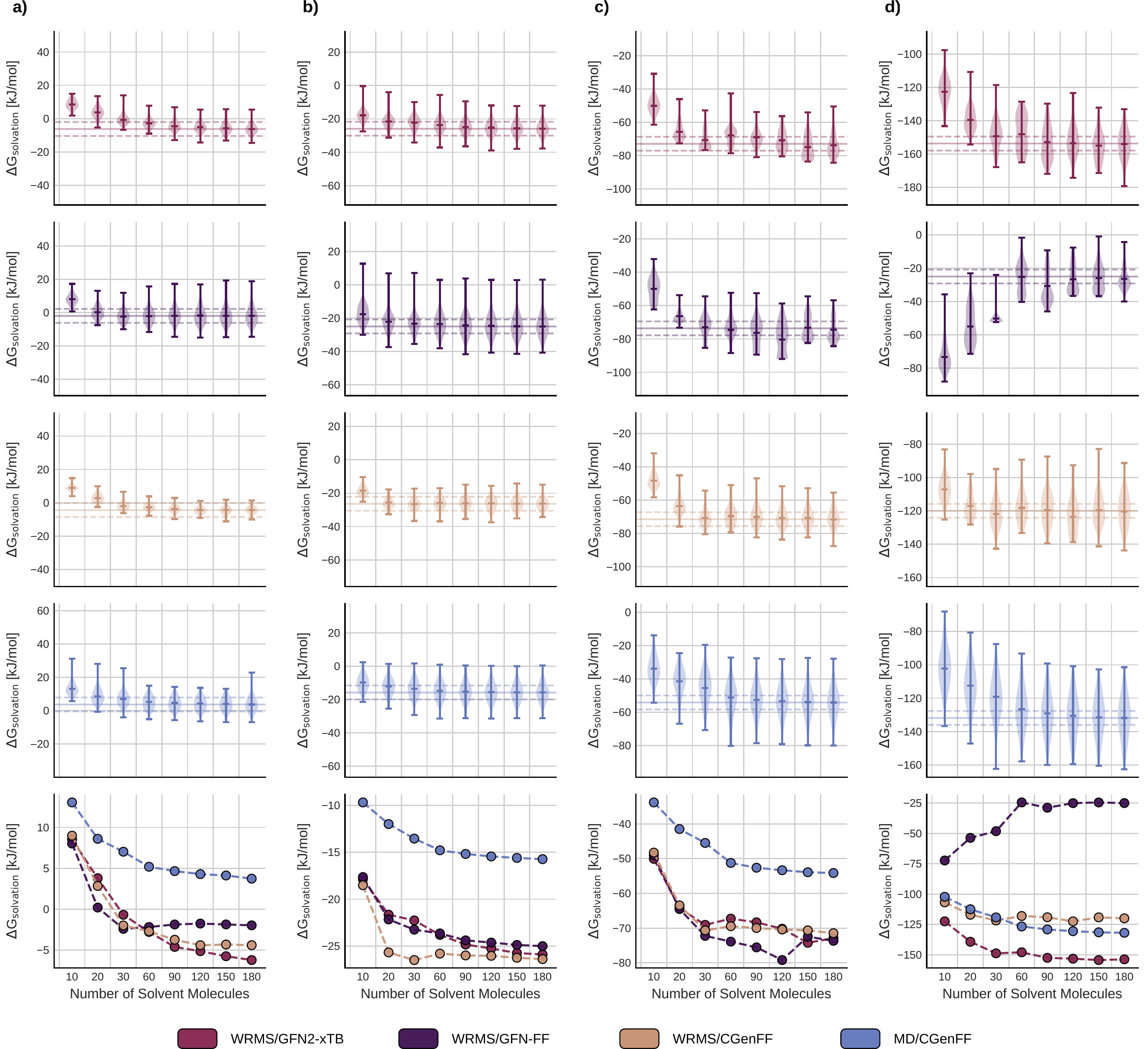}
    \caption{Solvation free energies for \ce{CH3Cl} (a), \ce{NH3} (b), the transition state (c) and the product (d) of the Menshutkin reaction in water derived with the four different combinations of sampling strategy and method (bottom). The first four rows show the distribution for each evaluated combination, indicated by color, the fifth row represents the direct comparison between the four approaches.}
    \label{fig:MenshutkinReaction_Comparsion_solvationEnergies}
\end{figure}

\begin{figure}[ht]
    \centering
    \includegraphics[width = 0.95\textwidth]{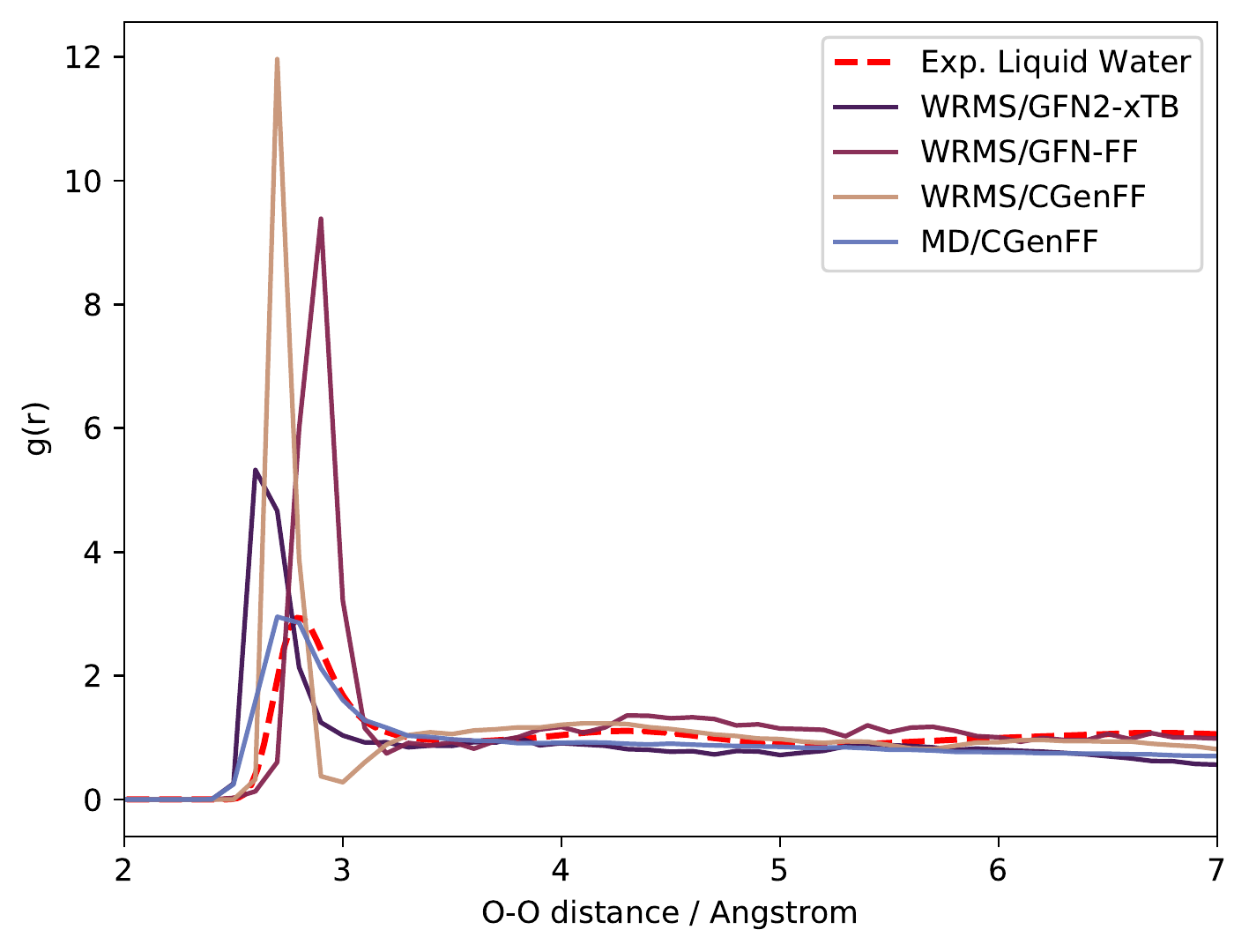}
    \caption{Radial distribution functions (RDF) of O-O distances in micosolvation water clusters. RDFs were generated for the transition state of Menshutkin reaction. All snapshots containing 180 water molecules for each sampling technique were used. Build-up data contains Boltzmann weights. Experimental data was taken from Ref~\citenum{Soper2000}.}
    \label{fig:rdf}
\end{figure}

For all species involved in the Menshutkin reaction, the target accuracy of $\pm$~\SI{4.18}{kJ / mol} ($\pm$~\SI{1}{kcal / mol}) around the mean of $\Delta {G_\mathrm{solvation}}$ with 180 water molecules is indicated in the upper four rows of Fig.~\ref{fig:MenshutkinReaction_Comparsion_solvationEnergies} to assess the convergence behaviour.
This window of accuracy is reached by all combinations studied.
The fastest convergence, in terms of added solvent molecules, can be observed with WRMS/CGenFF.
For both reagents and the TS, their $\Delta G_\mathrm{solvation}$ values obtained with any WRMS based combination converge to similar final energies whereas solvation free energies obtained with MD/CGenFF are systematically higher, about \SI{5}{kJ/mol} to \SI{20}{kJ/mol}.
For the solute--solvent clusters of the TS with 180 water molecules, the radial distribution function (RDF) of the \ce{O\bond{1}O} distances for all combinations and an experimental reference are shown in Fig.~\ref{fig:rdf}.
Here, any strategy with WRMS leads to curves with unrealistic peak heights while the MD snapshots result in a RDF curve with a peak close to the experimental reference.
Hence, WRMS are forming tighter clusters and upon closer inspection also more hydrogen bonds between water molecules than observed in experiment.
The effects of these and other structural differences on the free energy value are discussed below.

\begin{figure}
    \centering
    \includegraphics[width = 0.95\textwidth]{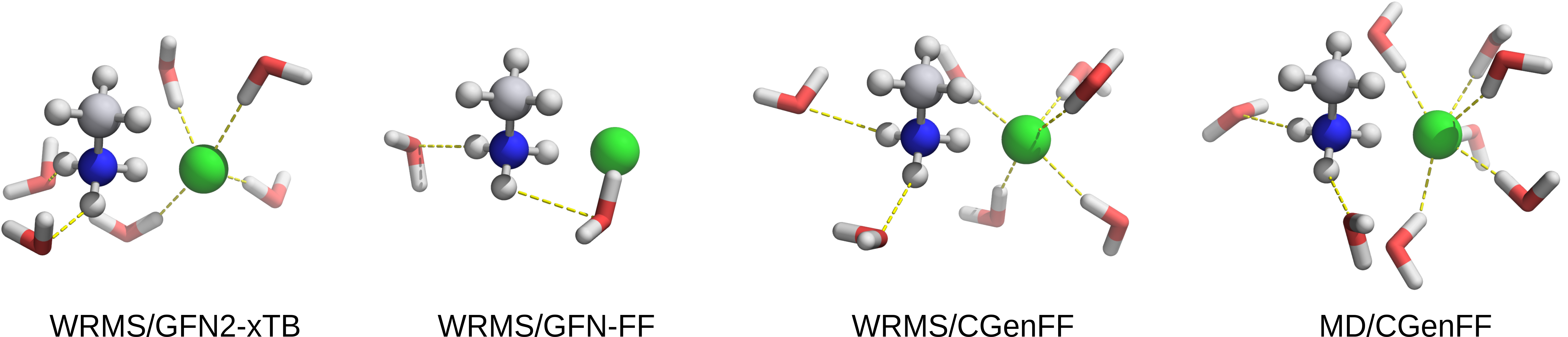}
    \caption{Product of the Menshutkin reaction with coordinating water molecules cut out from the solute--solvent cluster with 180 water molecules obtained with the four different combinations of sampling strategy and method. Coordination of water is indicated by yellow dashes.}
    \label{fig:Menshutkin_product_coordination}
\end{figure}

For the product, the convergence behaviour is different.
The solvation free energies with WRMS/GFN-FF converge in a concave instead of the usual convex way.
With WRMS/GFN2-xTB, the converged energies are significantly lower than with combinations of CGenFF.
With CGenFF, $\Delta G_\mathrm{solvation}$ energies from MD sampling are lower than from WRMS from 60 solvents onward.
Here, the dynamic sampling yields snapshots with higher solute stabilization than the static sampling despite both structure generations being based on CGenFF.
This is unique for the solutes examined in this work.
The partial charges in \ce{CH3(NH3)+\bond{...}Cl-}, the product of the Menshutkin reaction, are likely to cause the observations described above.
Especially the stabilization of \ce{Cl-} and \ce{NH3+} through hydrogen bonding with surrounding water molecules is responsible for low solvation free energies.
Analyzing the close coordination sphere of \ce{Cl-} and \ce{NH3+} of the structures with the highest Boltzman weight of the three differently obtained WRMS with 180 solvents and the structure with 180 solvents with the lowest solvation free energy of the MD sampling confirms this theory.
Excerpts of these structures, indicating solvent coordination are shown in Fig.~\ref{fig:Menshutkin_product_coordination}.

The structure of MD/CGenFF has the most and the shortest \ce{Cl^{-}\bond{...}H} bonds, followed by the structure of WRMS/CGenFF and of WRMS/GFN2-xTB.
The shorter bonds in the structure of MD/CGenFF compared to the one of WRMS/CGenFF are unexpected as one would expect the optimized structure of WRMS/CGenFF to show shorter bond lenghts.
However, the optimized structure has a higher number of \ce{OH\bond{...}O} bonds, according to the radial distribution functions (RDF) of \ce{O-O} distances shown in Fig.~\ref{fig:rdf}, which outweigh the benefit of short \ce{Cl^{-}\bond{...}H} bonds considering the whole solute--solvent cluster.
The structure of WRMS/GFN-FF shows no \ce{Cl^{-}\bond{...}H} bonds, the GFN-FF method prefers the formation of hydrogen bonds between water molecules rather than between the partially charged \ce{Cl^{-}} and water.
Hence, the concave curve of $\Delta G_\mathrm{solvation}$ determined with WRMS/GFN-FF for the product results.
All of the examined structures show two \ce{NH\bond{...}O} interactions.
Here, the structure of WRMS/GFN2-xTB shows short \ce{H-O} distances (\SI{1.82}{\angstrom} and \SI{1.89}{\angstrom}) which cause strong solute--solvent interactions when analyzing it with our sDFT hybrid model and lower the solvation free energy compared to WRMS/CGenFF.
Partial charges in the solute are clearly challenging for any method responsible for the calculation of energies and gradients during the structure optimization or the MD simulation and care must be taken when choosing a method, as the above excursion shows.

The distributions of $\Delta G_\mathrm{solvation}$ standing out, in terms of shape, for the TS and the product, with, \emph{e.g.}, 30 solvent molecules, obtained with WRMS/GFN-FF and WRMS/GFN2-xTB can be linked to insufficient sampling.
Here, structures within an ensemble with large Boltzman weights have more and shorter hydrogen bonds between solute and solvent than the majority of structures.
These special arrangements either break up with more added water molecules, in the case of WRMS/GFN-FF, or become more frequent, in the case of WRMS/GFN2-xTB, and hence lead to normal distributions again.
Increasing the sample size would yield distributions of normal shape as well.
However, as $\Delta G_\mathrm{solvation}$ is not converged where insufficient sampling occurs, we decided to prioritize convergence over an ideal distribution for pre-converged ensembles.

Of all the solutes analyzed in this work, solvation free energies obtained for \ce{CH3Cl} and \ce{NH3} can be compared to experimental reference data\cite{Matos2017, Mobley2018}, as shown in Tab.~\ref{tab:solvation_energies}.

\begin{table}[ht]
    \centering
    \begin{tabular}{l|rrrr|r}
          \toprule\toprule
          & \multicolumn{5}{c}{$\Delta G_\textrm{solvation}$ in kJ/mol} \\
                     & WRMS/GFN2-xTB & WRMS/GFN-FF & WRMS/CGenFF & MD/CGenFF & Exp.$^a$ \\ \midrule
          \ce{CH3Cl} & $-6.2(30)$ & $-2.0(52)$ & $-4.4(21)$ & $3.7(43)$ & $-2.3$ \\
          \ce{NH3}   & $-25.9(40)$ & $-25.0(63)$ & $-26.4(34)$ & $-15.7(55)$ & $-17.9$ \\
          \bottomrule\bottomrule
    \end{tabular}
    \caption{Solvation free energies ($\Delta G_\textrm{solvation}$) for chloromethane (\ce{CH3Cl}) and ammonia (\ce{NH3}). 
    Values taken from the data set with 180 water molecules.
    $^a$ Data published in Ref.~\citenum{Mobley2018}}
    \label{tab:solvation_energies}
\end{table}

The solvation free energies for ammonia are overestimated by all approaches with WRMS, only MD sampling returns energies for which the experimental reference is within the theoretical error of the method.
On the other hand, MD/CGenFF is unable to reproduce the experimental reference for chloromethane, whereas the experimental references lies within the error of WRMS/CGenFF and WRMS/GFN-FF.
With WRMS/GFN2-xTB, the solvation free energy for chloromethane is overestimated.
The tendency of underestimation of the magnitude of solvation free energies of chlorinated molecules in solvents considering different force fields in MD simulations has been noted recently.\cite{Kashefolgheta2020}\\

\begin{figure}[!ht]
    \centering
    \includegraphics[width = 0.73125\textwidth]{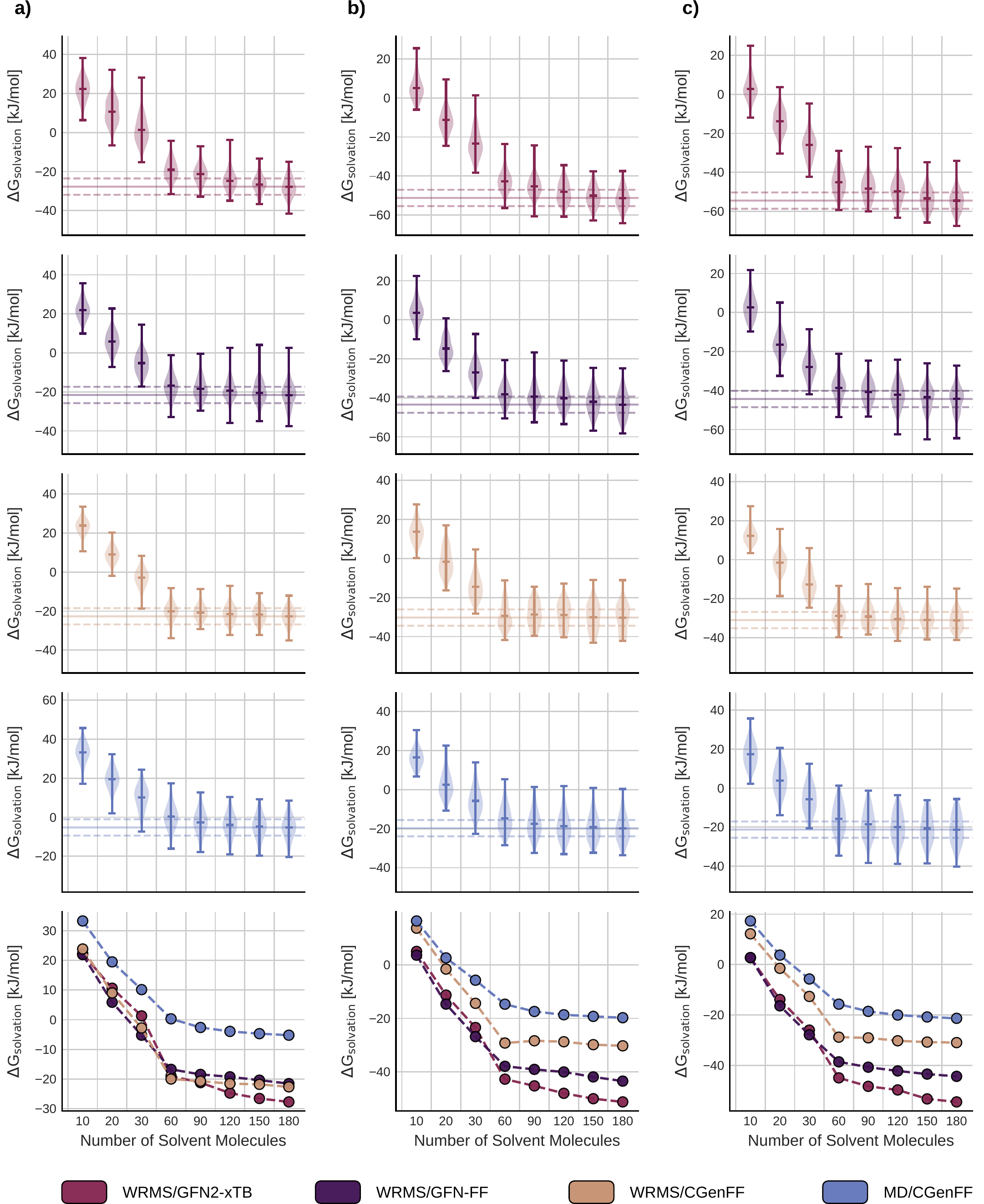}
    \caption{
    Solvation free energies for the reagent (a), the transition state (b) and the product (c) of the Claisen rearrangement in water derived with the four different combinations of sampling strategy and method (bottom). The first four rows show the distribution for each evaluated combination, indicated by color, the fifth row represents the direct comparison between the four approaches.
    }
    \label{fig:ClaisenRA_water_Comparsion_solvationEnergies}
\end{figure}

Instead of the intermolecular Menshutkin reaction, we now examine solvation free energies of species involved in the intramolecular Claisen rearrangement in two different solvents, starting with water.
As can be seen in Fig.~\ref{fig:ClaisenRA_water_Comparsion_solvationEnergies},  the target accuracy of $\pm$~\SI{4.18}{kJ / mol} ($\pm$~\SI{1}{kcal / mol}) around the mean of $\Delta {G_\mathrm{solvation}}$ with 180 water molecules is reached for all three species.
$\Delta G_\textrm{solvation}$ obtained with WRMS/GFN2-xTB converges in the slowest fashion, WRMS/CGenFF in the fastest where convergence is already reached with 60 water molecules for reagent, TS and product.
The solvation free energies of the product are lower than for the reagent.
This is due to the formation of the ketone group and resulting hydrogen bonds between the newly formed functional group and water.\cite{irani2009joint}
For the reagent, solvation free energies based on WRMS result in similar energies.
For the TS and the product, the solvation free energies of the strategies differ by about \SI{10}{kJ / mol} to the next lower energy.
The order of magnitude of the solvation free energy increases from MD/CGenFF to WRMS/CGenFF to WRMS/GFN-FF to WRMS/GFN2-xTB.
The product solvated with WRMS/GFN2-xTB experiences the larges stabilizing effect with increasing solvent number.
While the $\Delta G_\textrm{solvation}$ curves obtained with MD/CGenFF converge smoothly, the curves obtained with WRMS based methods display a significant kink from 30 to 60 solvent molecules, suggesting the completion of the solvation shell close to the solute.
The addition of solvent molecules from 60 solvent moleculs onward has less effect on the solvation free energy than the addition until a total of 60 solvent molecules.

\begin{figure}[!ht]
    \centering
    \includegraphics[width = 0.73125\textwidth]{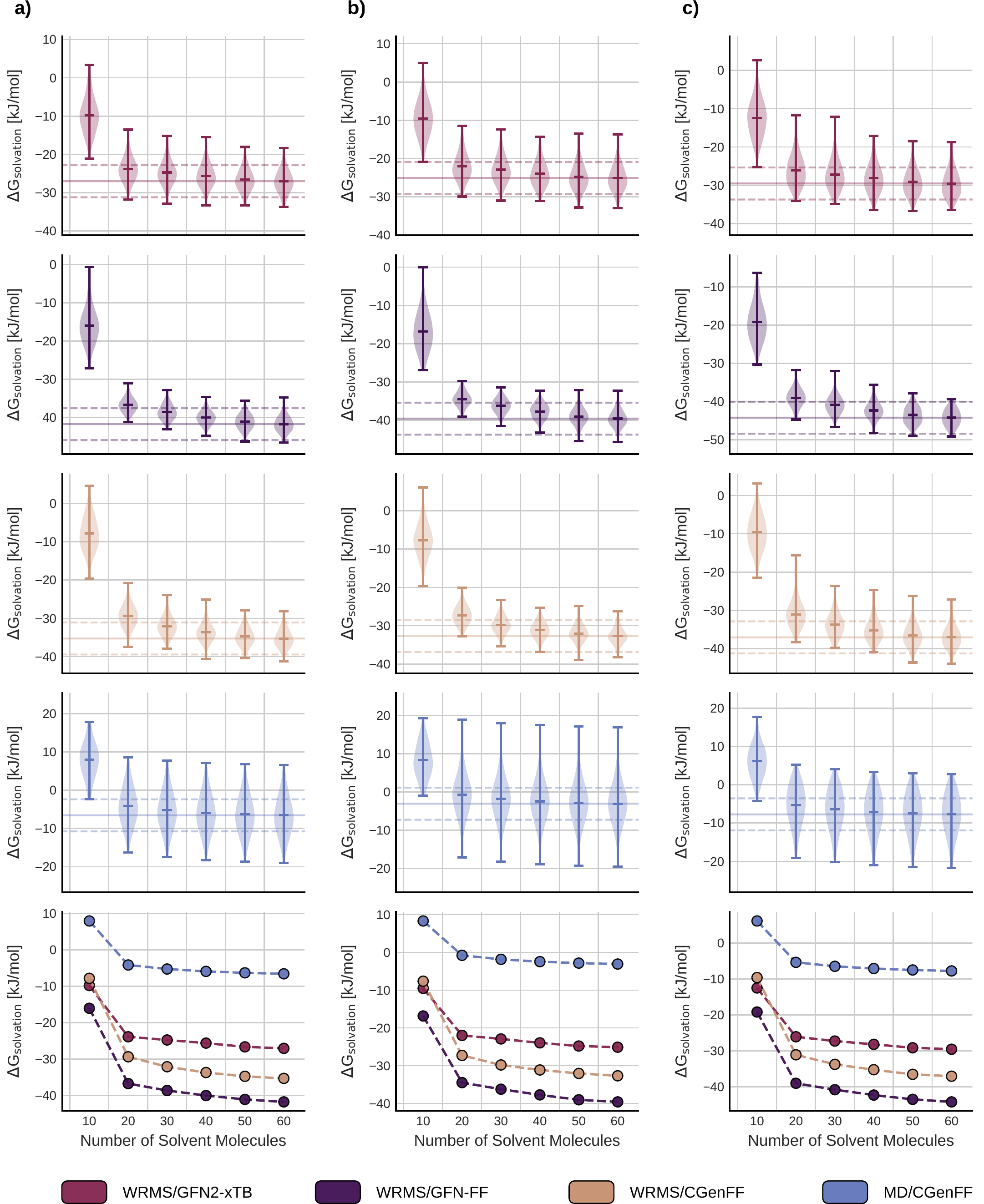}
    \caption{Solvation free energies for the reagent (a), the transition state (b) and the product (c) of the Claisen rearrangement in cyclohexane derived with the three different configuration sampling approaches (bottom). The first three rows show the distribution for each sampling approach, indicated by color, the fourth row represents the direct comparison between the three sampling approaches.}
    \label{fig:ClaisenRA_cyc_Comparsion_solvationEnergies}
\end{figure}

Additionally to the Claisen rearrangement in water, the solvation free energies of same reaction in cyclohexane are discussed here.
The solvation free energies for reagent, TS and product are shown in Fig.~\ref{fig:ClaisenRA_cyc_Comparsion_solvationEnergies}.
The target accuracy of $\pm$~\SI{4.18}{kJ / mol} ($\pm$~\SI{1}{kcal / mol}) around the mean of $\Delta {G_\mathrm{solvation}}$ with 60 cyclohexane molecules is reached at least with 30 solvent molecules for all strategies.
Solvation free energies obtained with the WRMS/GFN-FF combination result in the largest magnitude of $\Delta G_\textrm{solvation}$ for all species.
Furthermore, the order in terms of magnitude of solvation free energies is the same for the reagent, the TS and the product.
The magnitude increases from MD/CGenFF to WRMS/GFN2-xTB to WRMS/CGenFF to WRMS/GFN-FF.
The kink from 10 to 20 solvent molecules indicates the completion of a solvent shell as additional solvent molecules do not alter the solvation energy as much.
The strategies employing WRMS are more alike with each other in terms of solvation free energies for reagent, TS and product than with the MD/CGenFF approach.
In contrast to the Claisen rearrangement in water, the TS and the product are not stabilized as much in cyclohexane as in the polar solvent.
For the reagent, the stabilizing effects are stronger in cyclohexane than in water when employing WRMS/CGenFF and, especially, WRMS/GFN-FF.
As discussed previously, this might be linked to the preference of GFN-FF to form hydrogen bonds between different water molecules rather than between water molecules and the solute.
WRMS*GFN2-xTB and MD/CGenFFshow similar stabilizing effects for the reagent in water as for the reagent in cyclohexane.

\subsection{Comparison with Experimental Results}

The effects of solvation are more pronounced when studying activation barriers and kinetics and the reaction free energies.
Starting with the theoretical relative energies of the Menshutkin reaction in the gas phase, the free activation barrier is \SI{193.9}{kJ/mol} and the reaction itself slightly endothermic.
We determined the barrier for the same reaction with iodomethane instead of chloromethane to be \SI{102(6)}{kJ/mol} at \SI{25}{\degree C} by linear regression of the Eyring equation from the experimental data published in Ref.~\citenum{Okamoto1967}.
This serves as the lower estimate for the free energy of activation, assuming iodide being a better leaving group than chloride in water, as explained above.

\begin{table}[ht]
    \centering
    \begin{tabular}{l|rr}
          \toprule\toprule
          & $\Delta^{\ddagger}$G & $\Delta_{R}$G \\ \midrule
          & \multicolumn{2}{c}{in kJ/mol} \\ \midrule
          Gas Phase (Electronic Energy Only)  & $143.5$   & $-39.4$ \\
          Gas Phase                           & $193.9$   &   $2.5$ \\ \midrule
          Water (CPCM)                        & $105.3$   & $-47.7$ \\
          Water (COSMO-RS)                    & $103.2$   & $-75.8$ \\ \midrule
          Water (WRMS/GFN2-xTB)$^b$           & $107.1$   & $-69.8$ \\
          Water (WRMS/GFN-FF)$^b$             & $101.2$   &  $53.8$ \\
          Water (WRMS/CGenFF)$^b$             & $107.2$   & $-37.3$ \\
          Water (MD/CGenFF)$^b$               & $105.7$   & $-68.0$ \\ \midrule
          Water (Exp.)$^a$                    & $> 101.7$ & -       \\
          \bottomrule\bottomrule
    \end{tabular}
    \caption{Gibbs energies for the reaction ($\Delta_{R}$G) and 
             activation ($\Delta^{\ddagger}$G) of the Menshutkin reaction. 
             Values tabulated for methods that explicitly treat the solvent 
             were taken from the data set with the most solvent molecules 
             showed in the previous sections. $^a$ Values obtained by linear regression of the Eyring equation from the data published in Ref.~\citenum{Okamoto1967}.
             $^b$Data generated using sDFT as described in this work.}
    \label{tab:menshutkin_results}
\end{table}

\begin{figure}[!ht]
    \centering
    \includegraphics[width = 0.73125\textwidth]{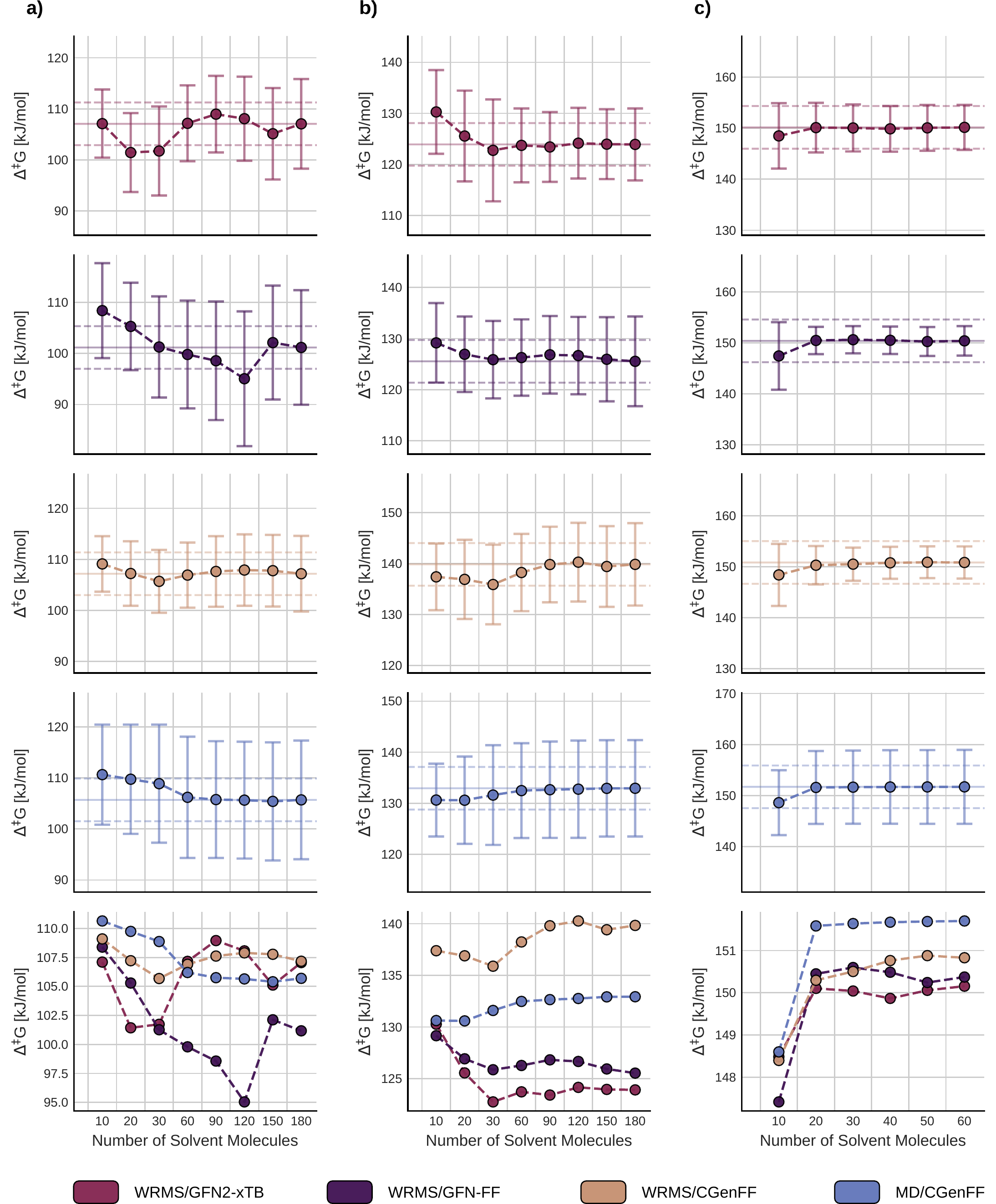}
    \caption{Free energies of activation for the Menshutkin reaction (a), the Claisen rearrangement in water (b) and in cyclohexane (c) derived with the four different combinations of sampling strategy and method. The first four rows show the convergence and error as well as a \SI{4.18}{kJ / mol} window, the fifth row represents the direct comparison between the four investigated strategies.}
    \label{fig:Reactioncomparsion_barriers}
\end{figure}

All results concerning the Menshutkin reaction are summarized in Tab.~\ref{tab:menshutkin_results}.
The approaches based on CPCM\cite{Barone1998} and COSMO-RS\cite{Klamt2017} as well as on MD/CGenFF are slightly above the lower estimate.
About \SI{2}{kJ/mol} higher than the previous three are the barriers obtained with WRMS/GFN2-xTB and WRMS/CGenFF.
The barrier resulting from WRMS/GFN-FF is lower than the lower estimate.
The combination based on GFN-FF seems to be insufficient to capture the experimental trend.
Furthermore, the curve of $\Delta^{\ddagger} G$ obtained with WRMS/GFN-FF features an unexpected step from 120 to 150 solvent molecules, as can be seen in Fig.~\ref{fig:Reactioncomparsion_barriers}a).
In contrast, the curves obtained with methods employing CGenFF, regardless of the sampling approach, converge steadily.
The wiggles in the curve based on WRMS/GFN2-xTB can be related to the wiggles of $\Delta G_\textrm{solvation}$ of the TS.

Upon solvation, the reaction becomes exothermic due to the stabilization of the product through water.
The magnitude of the reaction free energy obtained with COSMO-RS is the largest exothermic one of all methods studied, employing WRMS/CGenFF results in $\Delta_\text{R}G$ with the smallest magnitude of exothermic reaction free energies.
The reaction free energy obtained with WRMS/GFN-FF is endothermic due to the lack of product stabilization through water, as discussed previously.

\begin{figure}[!ht]
    \centering
    \includegraphics[width = 0.73125\textwidth]{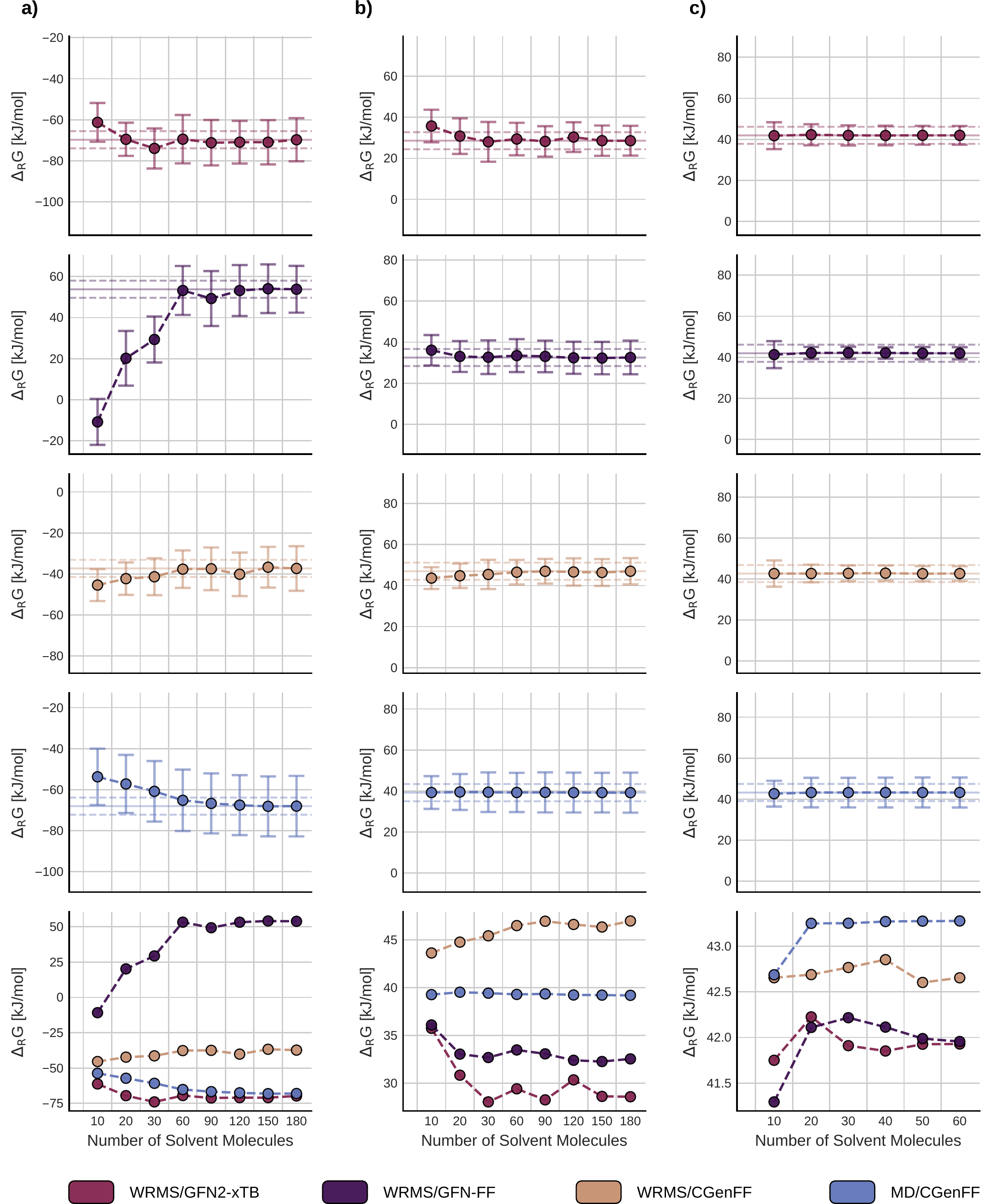}
    \caption{Free energies for the Menshutkin reaction (a), the Claisen Rearrangement in water (b) and in cyclohexane (c) derived with with the four different combinations of sampling strategy and method. The first four rows show the convergence and error as well as a \SI{4.18}{kJ / mol} window, the fifth row represents the direct comparison between the four sampling approaches.}
    \label{fig:Reactioncomparsion_reactionEnergy}
\end{figure}
The analysis of the barriers of the Menshutkin reaction shows that three of four examined strategies with our sDFT hybrid model are able to capture the experimental trend.
Even the estimate resulting from WRMS/GFN-FF is not significantly below the experimental lower estimate.\\

In the next model system of this work, the Claisen rearrangement, we examine the influence of different solvents on the same reaction.
For the Claisen rearrangement in water and cyclohexane, experimental reference provides more than an estimate of the lower barrier.\cite{White1970}
The extensive experimental study derives $\Delta^{\ddagger} H$ and $\Delta^{\ddagger} S$ values for the reaction in the gas phase, the reaction in an ethanol-water mixture (reference system for our reaction in water) and the reaction in tetradecane (reference system for our reaction in cyclohexane) in a temperature range between \SI{150}{\degree C} and \SI{230}{\degree C}.
Assuming $\Delta^{\ddagger} H$ and $\Delta^{\ddagger} S$ are temperature independent, estimates for $\Delta^{\ddagger} G$ at \SI{25}{\degree C} are obtained via extrapolation allowing a direct comparison between the barriers obtained with our hybrid sDFT model and the barriers derived from the described experimental model systems.
To capture the acceleration of the reaction upon solvation, we introduce the shift in the free energy of activation upon solvation $\Delta\Delta^{\ddagger} G$,
defined as $\Delta^{\ddagger} G_\text{solvent} - \Delta^{\ddagger} G_\text{gas phase}$.
Our results as well as the results based on the experimental reference systems are summarized in Tab.~\ref{tab:claisen_results}.

\begin{table}[ht]
    \centering
    \begin{tabular}{l|rr|r}
        \toprule\toprule
          &$\Delta^{\ddagger}$G & $\Delta_{R}$G & $\Delta\Delta^{\ddagger}$G \\ \midrule
          & \multicolumn{3}{c}{in kJ/mol} \\ \midrule
          Gas Phase (Electronic Energy Only) & $147.0$ & $45.8$ &              \\
          Gas Phase                          & $149.3$ & $44.9$ &              \\ 
          Gas Phase (Exp.)$^b$     & $156.5$ & -      &              \\ \midrule
          Water (CPCM)                       & $137.4$ & $42.2$ & $-11.9$      \\
          Water (COSMO-RS)                   & $130.3$ & $35.5$ & $-19.0$      \\
          Water (WRMS/GFN2-xTB)$^a$          & $123.9$ & $28.6$ & $-25.4$     \\ 
          Water (WRMS/GFN-FF)$^a$            & $125.5$ & $32.5$ & $-23.8$      \\
          Water (WRMS/CGenFF)$^a$            & $139.8$ & $47.0$ & $-9.5$      \\
          Water (MD/CGenFF)$^a$              & $132.9$ & $39.2$ & $-16.4$     \\ \midrule
          $28.5\%$ Ethanol-Water (Exp.)$^b$  & $132.7$ & - & $-23.8$ \\ 
          \midrule
          Cyclohexane (CPCM)                 & $146.1$ & $39.4$ & $-3.2$ \\
          Cyclohexane (COSMO-RS)             & $147.8$ & $41.4$ & $-1.5$ \\
          Cyclohexane (WRMS/GFN2-xTB)$^a$    & $150.2$ & $41.9$ & $0.9$  \\
          Cyclohexane (WRMS/GFN-FF)$^a$      & $150.4$ & $42.0$ & $1.1$  \\
          Cyclohexane (WRMS/CGenFF)$^a$      & $150.8$ & $42.7$ & $1.5$  \\
          Cyclohexane (MD/CGenFF)$^a$        & $151.7$ & $43.3$ & $2.4$  \\ \midrule
          Tetradecane (Exp.)$^b$             & $150.7$ & -      & $-5.8$ \\
          \bottomrule\bottomrule
    \end{tabular}
    \caption{Gibbs free energies for the reaction ($\Delta_{R}$G) and 
             activation ($\Delta^{\ddagger}$G) of the Claisen rearrangement. 
             Values tabulated for methods that explicitly treat the solvent 
             were taken from the data set with the most solvent molecules
             shown in the previous sections. $^a$Data generated using sDFT 
             as described in this work. $^b$Values obtained  by  extrapolation to \SI{25}{\degree C} employing the  data  published  in  Ref.\citenum{White1970}.}
    \label{tab:claisen_results}
\end{table}

The theoretical value of the barrier in the gas phase does not match the experimental reference.
However, as the gas phase barriers serve as relative zero points for the experiment as well as for the different examined strategies, in the context of this work an exact match is not required.
The theoretical reaction free energy in the gas phase is endothermic with $\Delta_\textrm{R} G=44.9$~\si{kJ/mol}.
Recall, however, that the Claisen rearrangement is driven by the tautomerization after the rearrangement which is not further studied in this work.
The experimental result states an acceleration of the reaction in  the ethanol-water mixture, as given by $\Delta\Delta^{\ddagger}G=-23.8$~\si{kJ/mol}.
The acceleration resulting from the implicit solvent models are lower than the experimental reference.
With CPCM, the barrier is higher than with COSMO-RS.
Compared to the Menshutkin reaction, the difference in the results of the implicit models is larger.
COSMO-RS is matching the experimental result better.
As can be seen in Fig.~\ref{fig:Reactioncomparsion_barriers}b), all $\Delta ^{\ddagger}G$ curves converge and the final value with 180 water molecules deviates from the start values with 10 water molecules by about \SI{5}{kJ/mol}.
The experimental acceleration is captured with WRMS/GFN-FF, WRMS/GFN2-xTB and MD/CGenFF.
The acceleration predicted by WRMS/CGenFF is significantly lower than the experimental estimate.
Considering $\Delta_\textrm{R} G$, the implicit solvent models lead to a less endothermic reaction compared to the gas phase.
Here, COSMOS-RS is stabilizing the product better than CPCM, showing the same trend as for the activation barrier.
The curves of $\Delta_\textrm{R} G$ do not vary greatly with increasing number of solvent molecules, as shown in Fig.~\ref{fig:Reactioncomparsion_reactionEnergy}b).
The reaction has the lowest $\Delta_\textrm{R} G$ energy with WRMS/GFN2-xTB followed by WRMS/GFN-FF.
With MD/CGenFF, the energy is between the one obtained with CPCM and COSMO-RS.
Surprisingly, with WRMS/CGenFF the result for $\Delta_\textrm{R} G$ is higher than for the gas phase, however not significantly. 

Switching from the polar to the unpolar solvent, an acceleration of the reaction, compared to the gas phase, can be experimentally observed in tetradecane as well.
With $\Delta\Delta^{\ddagger} G=-5.8$~\si{kJ/mol}, the acceleration is clearly less pronounced than in water.
The implicit solvent models are both able to capture this effect, CPCM slightly better than COSMO-RS.
The curves of $\Delta^{\ddagger} G$ converge in a concave fashion for all strategies, shown in Fig.~\ref{fig:Reactioncomparsion_barriers}c).
The four strategies employing our sDFT hybrid model all suggest a higher barrier in cyclohexane than in the gas phase.
However, the small accelerating effect observed in experiment is still within the
statistical error of these strategies.
For the $\Delta_\textrm{R} G$ data, which are shown in Fig.~\ref{fig:Reactioncomparsion_reactionEnergy}c), the values from 10 to 60 cyclohexane molecules do not change more than about \SI{1}{kJ/mol} for all strategies.
All strategies, including the implicit solvents, lower the reaction free energy compared to the gas phase.

\section{Conclusions}
In this work, we introduced a hybrid model that incorporates explicit solvent molecules into the polarizable continuum model formalism using subsystem density functional theory.
Given the dependence on the configuration of the explicit solvent molecules in the model, two different sampling techniques and three different quantum chemical methods were employed to analyze the new model.
With proper sampling the sDFT-hybrid model is accurate when modeling the kinetic descriptors of two literature-known experiments.
The quality of the results depends on the chosen sampling method and also on the molecular model.
While the sDFT-hybrid model is of course computationally more costly than purely implicit solvation, we have focused on two sampling approaches that can generate solvation data in relatively short time.
One relies on the sampling of solvent structures using molecular dynamics with standard force fields, the other optimizes static snapshots with (semi-empirical) quantum chemistry methods and subsequently weighs them in accordance with a given temperature.\\
The molecular dynamics sampled data is more reliable than the data based on optimized structures.
This is best illustrated by the O--O radial distribution function of the solvent water generated from the sampled examples.\\
One drawback of the current MD approach is that it is not easily automatized when new solvents and solute molecules require new parameter sets.
To this end, molecular dynamics approaches based on fast, computationally inexpensive semi-empirical methods\cite{Husch2018} or automatically generated system focused force fields\cite{Brunken2020, Brunken2021} may be the next logical step.
Additionally, it would be beneficial to allow extra- or interpolation for temperature regimes based on a fixed set of MD trajectories, such that slight changes in the temperature do not require reruns of these MD simulations.\\
Concerning the coupling of PCM and sDFT, future work should explore the calculation of solute and solvent Hessians within the sDFT framework, allowing for a better description of the thermal motion contributions to the solvation free energy and removing auxiliary PCM-only calculations that are currently required.
It is obvious that recent developments in machine learning for chemical applications could speed up the sampling of the mostly organic solvent molecule complexes\cite{Eastman2017, Behler2017}. 

\begin{acknowledgement}
This work was financially supported by the Deutsche Forschungsgemeinschaft (DFG) (GZ: UN 417/1-1 and SFB858 project Z01), by the ETH Research Grant ETH-44 20-1, and by the Schweizerischer Nationalfonds (SNF) (Project 200021\_182400).

\end{acknowledgement}

\providecommand{\latin}[1]{#1}
\makeatletter
\providecommand{\doi}
  {\begingroup\let\do\@makeother\dospecials
  \catcode`\{=1 \catcode`\}=2 \doi@aux}
\providecommand{\doi@aux}[1]{\endgroup\texttt{#1}}
\makeatother
\providecommand*\mcitethebibliography{\thebibliography}
\csname @ifundefined\endcsname{endmcitethebibliography}
  {\let\endmcitethebibliography\endthebibliography}{}

\end{document}